# The prominent and heterogeneous gender disparities in scientific novelty: evidence from biomedical doctoral theses

Meijun Liu[1,*], Zihan Xie[2], Alex Jie Yang[3], Chao Yu[4], Jian Xu[4], Ying Ding[5], Yi Bu[6],

[1] Institute for Global Public Policy, Fudan University, Shanghai, China
[2] School of Management, Fudan University, Shanghai, China
[3] School of Information Management, Nanjing University, Nanjing, China
[4] School of Information Management, Sun Yat-sen University, Guangzhou, China
[5] School of Information, University of Texas at Austin, Austin, TX, USA
[6] Department of Information Management, Peking University, Beijing, China

* Corresponding author
Email address: meijunliu@fudan.edu.cn (M. Liu)

**Abstract**
Scientific novelty is the essential driving force for research breakthroughs and innovation. However, little is known about how early-career scientists pursue novel research paths, and the gender disparities in this process. To address this research gap, this study investigates a comprehensive dataset of 279,424 doctoral theses in biomedical sciences authored by US Ph.D. graduates. Spanning from 1980 to 2016, the data originates from the ProQuest Dissertations & Theses Database. This study aims to shed light on Ph.D. students' pursuit of scientific novelty in their doctoral theses and assess gender-related differences in this process. Using a combinatorial approach and a pre-trained Bio-BERT model, we quantify the scientific novelty of doctoral theses based on bio-entities. Applying fractional logistic and quantile regression models, this study reveals a decreasing trend in scientific novelty over time and heterogeneous gender disparities in doctoral theses. Specifically, female students consistently exhibited lower scientific novelty levels than their male peers. When supervised by female advisors, students' theses are found to be less novel than those under male advisors. The significant interaction effect of female students and female advisors suggests that female advisors may amplify the gender disparity in scientific novelty. Moreover, heterogeneous gender disparities in scientific novelty are identified, with non-top-tier universities displaying more pronounced disparities, while the differences at higher percentile ranges were comparatively more minor. These findings indicate a potential underrepresentation of female scientists pursuing novel research during the early stages of their careers. Notably, the outcomes of this study hold significant policy implications for advancing the careers of female scientists.

# 1 Introduction

Novel research plays a significant role as a catalyst for scientific breakthroughs and technological innovation. It possesses the potential to become a breakthrough on its own and trigger subsequent advancements that may have far-reaching impacts (Criscuolo et al., 2017). Drawing on Schumpeter's perspective on innovation, it has been widely recognized that innovation emerges through the novel combination of existing knowledge (Fleming, 2001). Given its importance, a growing emphasis is placed on pursuing scientific novelty (Cohen, 2017). Early-career scientists are encouraged to generate novel knowledge and bring new ideas to science, as evidenced by the establishment of initiatives such as the NIH Director's New Innovator Award.[1]

Although there is a substantial body of literature on measuring scientific novelty (Liu, Bu, et al., 2022; Uzzi et al., 2013) and investigating its origins and development, limited research has focused on how early-career scientists pursue novel research. Similarly, there is a lack of exploration regarding gender disparities within this context. Scientific novelty presents a paradoxical dual implication. On the one hand, pursuing novel research holds advantages for the career advancement of early-career scientists as it may lead to significant returns in the future. However, highly innovative ideas often carry inherent risks and can be vulnerable to recognition biases and delays (Trapido, 2015; Wang et al., 2017). In the face of the "publish or perish" imperative, early-career scientists face challenges balancing their pursuit of novel research with the need to publish their findings consistently. Understanding how early-career scientists produce novel scientific work is crucial to determine whether their research strategies align with the increasing value placed on groundbreaking knowledge in the scientific community, and to identify the mechanisms that facilitate a successful academic career.

Gender disparities in science have persisted throughout history and continue to exist today. STEM fields, in particular, are often perceived as male-dominated, and gender stereotypes about the roles and abilities of female scientists remain deeply ingrained within academia (Carli et al., 2016; Eagly et al., 2020). Unfortunately, the current research system inadequately supports female scientists, as they face disadvantages in accessing research funding (Larivière et al., 2011), scientific collaboration and leadership opportunities (Liu, Zhang, et al., 2022), and the challenges of balancing childcare responsibilities with career advancement (Fox, 2005). Empirical evidence from previous studies consistently demonstrates persistent gender gaps in science, including higher attrition rates, shorter academic careers, lower productivity, and diminished impact among female scientists relative to their male counterparts (Huang et al., 2020). Additionally, the underrepresentation of women in science results in a scarcity of role models, undermining women's confidence and sense of belonging in the field (Breda et al., 2023).

These aforementioned gender-related disparities can manifest in scientific novelty within academia. Female and male students may exhibit differences in pursuing novel research in their doctoral theses. The Matilda effect, which refers to the tendency for women's abilities and contributions to be underestimated and undervalued (Rossiter, 1993), is especially relevant to novel research. Based on role congruity theory, innovative work is often associated with masculine traits and is more commonly attributed to men than women (Luksyte et al., 2018). Novel research is expected to be produced by male scientists to a greater extent than their female counterparts. When women produce novel work, they are perceived as deviating from gender stereotypes. As a result of this sex-based stereotype, novel research generated by female scientists may not be evaluated or rewarded to the same degree as that demonstrated by their male colleagues. Empirical evidence suggests that women receive less recognition than men for their innovative contributions in various domains, including academia (Kabat-Farr & Cortina, 2012; Schmutz & Faupel, 2010; Trapido, 2022). Due to this gender-related penalty for novelty, female scientists may be more hesitant to pursue novel research.

---

[1] https://researchtraining.nih.gov/programs/other-training-related/DP2

Gender disparities in novel research can exhibit heterogeneities based on factors such as university prestige and the distribution of scientific novelty. These disparities may differ across institutions with varying levels of prestige. Empirical studies have shown that female scientists often face more significant disadvantages in elite universities, as evidenced by lower enrollment numbers, fewer opportunities for promotion (Jacobs, 1996), and limited access to leadership positions. However, it is worth noting that the work environment in top-tier universities may be more accommodating for female scientists, as these institutions are increasingly committed to promoting gender equity beyond mere non-discrimination measures (Bothwell et al., 2022). Consequently, gender disparities in scientific novelty may be less pronounced in top-tier universities. Furthermore, gender disparities in novel research may also vary between non-high performers and high performers in terms of the level of novelty in scientific work. Previous research suggests differences in gender disparity regarding publications and citations exist, highlighting that the underrepresentation of female scientists among scientific elites may be less significant, and they may even outperform their male counterparts (Chan & Torgler, 2020).

Based on 279,424 doctoral theses in biomedical sciences from US institutions from 1980 to 2016, this study investigates temporal and gender disparities in scientific novelty within doctoral theses and the potential heterogeneities concerning such gender disparities. This study makes both significant theoretical and practical contributions. The findings reveal the salient gender differences in research strategies concerning producing novel research during the formative stages of scientists' academic careers. Second, the findings enrich our understanding of the underlying mechanisms behind phenomena such as the "productivity puzzle" or "glass ceiling", shedding light on gender gaps in the biomedical workforce. Additionally, this study provides insights into the influence of gender on the development of scientific novelty. From the practical perspective, the measurement of scientific novelty this study applies can be extended to assess the scientific novelty of doctoral theses in disciplines that are similar to biomedical sciences and that of early-career scientists. The method to measure the scientific novelty of doctoral theses offers a nuanced evaluation of early-career scientists' innovation potential to funding agencies, institutions' managers and policymakers, which facilitates their decision-making processes regarding funding allocation, academic hiring, and promotions.

## 2 Research objectives

To uncover how early-career scientists pursue novel research, especially the gender disparity in this process, as well as its potential heterogeneities, this study proposes three research questions:

RQ1: How has the scientific novelty of doctoral theses evolved over the past decades?

RQ2: Are there any gender disparities in the scientific novelty of doctoral theses?

RQ3: Do gender disparities in the scientific novelty of doctoral theses exhibit heterogeneities?

In addressing the above research questions, we take the following steps:

(1) We adopt the combinatorial perspective of novelty to assess scientific novelty scores in biomedicine. Specifically, we measure these scores based on fine-grained knowledge units extracted from biomedical literature, namely bio-entities derived from titles and abstracts of doctoral theses. We further employ a pre-trained Bio-BERT model, trained on 29 million PubMed articles, to capture the semantic relationships between different bio-entities. To determine novelty, we classify a pair of bio-entities as novel if they exhibit rare combinations. The scientific novelty score of each doctoral thesis is then quantified as the fraction of novel entity pairs relative to the total number of entity pairs within the thesis.

(2) We utilize Welch's t-tests and linear regression analyses to explore the temporal pattern of scientific novelty in doctoral theses over the previous decades. Additionally, we assess whether the changes in scientific novelty are consistent across different categories, including university prestige and students' gender.

(3) We employ logistic and fractional regression models to analyze gender disparities in the scientific novelty of doctoral theses. Controlling for various influential factors, we compare the differences based on students' and advisors' genders. To shed light on whether the gender combination of students and advisors is related to the scientific novelty of doctoral theses, we examine an interaction effect of students' and advisors' gender on this aspect.

(4) To uncover potential variations in gender disparities that exist in scientific novelty among doctoral theses, we employ a two-step methodology. Firstly, subgroup regression analyses are utilized to examine whether there are differences in the extent of gender disparities in scientific novelty between top-tier and non-top-tier universities. Secondly, quantile regression analyses are carried out to explore whether gender disparities in scientific novelty vary across different quantiles.

The remainder of this study is organized as follows. The next section reviews the related work and proposes the research questions. Section 4 introduces the details of data and empirical approaches. In section 5, the results are presented. The last section discusses the findings and implications for science policies.

## 3 Literature review and research questions

In line with each research question, this study conducts a comprehensive review of the latest literature regarding the temporal evolution of scientific novelty, gender disparities in scientific novelty, and the heterogeneous nature of these disparities.

### 3.1 Temporal evolution of scientific novelty

The existing literature provides contrasting opinions of how scientific novelty developed. The literature on this topic can be broadly categorized into two streams: one focusing on the negative effect of the "knowledge burden" phenomenon and the other emphasizing the promoting effect of emerging techniques and the increasing need for novel solutions that may accelerate innovative processes.

Jones (2009) presents the concept of the "knowledge burden" mechanism, which offers insights into slowing innovative activities. According to this perspective, as the burden of knowledge grows, the value of new knowledge needed to compensate also increases, demanding higher-quality innovations. As knowledge expands, subsequent generations of researchers face a growing educational burden and are compelled to narrow their areas of expertise. Consequently, their capacities become limited, making innovation more challenging over time. Previous studies provide empirical evidence that aligns with this view of a declining trend in scientific novelty. For example, there has been a rise in the age at which inventors first patent and a general increase in the size of research teams that are used to compensate for innovators' reduced skills (Jones, 2010). Bloom et al. (2020) present empirical evidence from various industries, suggesting a significant increase in research efforts, but a sharp decrease in research productivity, and concluding that new ideas are getting harder and harder to find. Based on a large-scale dataset of publications and patents, Park et al. (2023) apply a new indicator of innovative activities, i.e., the CD index, to reflect the disruptive nature of innovation, and observe the declining disruptiveness of papers and patents over time.

The second stream of literature focuses on the positive impact of accumulated knowledge, emerging techniques, and the growing necessity to tackle complex research problems, which all contribute to accelerating the innovative process. As Newton stated, "If I have seen further, it is by standing on the shoulders of giants", indicating that knowledge begets new knowledge. The accumulation of knowledge creates a foundation conducive to scientific novelty, making pursuing novel research potentially easier over time. Many researchers argue that rapid advancements in emerging technologies such as artificial intelligence, crowdsourcing, and large language models (LLMs) can accelerate innovation (Kittur et al., 2019). These technologies serve as "general purpose inventions in the method of invention" that enhance the efficiency of conducting novel research (Cockburn et al., 2018). In the digital age, knowledge is being encoded at a large scale, and artificial intelligence (AI) can identify related concepts iteratively and automatically with researchers, thus accelerating the process of producing novel knowledge (Dwivedi et al., 2023). Furthermore, the unprecedented need for urgent and innovative solutions to complex global problems also contributes to the potential acceleration of novel research from the demand side. For instance, Liu, Bu, et al. (2022) analyzed nearly 100,000 papers related to the COVID-19 pandemic and observed a sudden increase in scientific novelty during this period. This finding suggests that, with the increasing globalization and complexity of socio-economic issues, scientific novelty is likely to rise substantially to address complex global challenges.

However, the existing literature has not adequately explored the developmental process of scientific novelty in doctoral theses. Furthermore, predictions based on existing theories yield contradictory answers regarding this question.

## 3.2 The gender disparity in scientific novelty and its heterogeneity

Extensive research has been conducted to investigate the underrepresentation of women in science, manifested in various forms. These include higher dropout rates, lower productivity (Aguinis et al., 2018), reduced scientific impact and visibility (Nittrouer et al., 2018), shorter research careers (Huang et al., 2020), limited access to senior authorship positions and prestigious journals (Holman et al., 2018), fewer opportunities for grants (Larivière et al., 2011), fewer awards (Ma et al., 2019), lower salaries (Hopkins, 2002), and underrepresentation in leadership positions within academia (Liu et al., 2023). Some studies have specifically investigated gender gaps in publication rates and citations among early-career students, including doctoral students (Lubienski et al., 2018; Schaller, 2022). While there is no consensus on the exact reasons behind these gender gaps in science, numerous factors have been considered contributors to the multifaceted disadvantage female scientists face. These factors range from cultural barriers (Reuben et al., 2014), and differential career choices (Ceci & Williams, 2011) to challenges related to childbearing and family responsibilities (Shen, 2013), sexual harassment (Karami et al., 2020), unequal allocation of research resources, and sex discrimination in grant and manuscript reviewing (Bornmann et al., 2007; Borsuk et al., 2009; Budden et al., 2008), as well as hiring and promotion practices (Sheltzer & Smith, 2014; Way et al., 2016).

Limited research exists on gender differences in research strategies, particularly concerning the pursuit of novel research. Insufficient focus has been directed toward comprehending the gender disparities in pursuing novel work, resulting in predominant normative discussions regarding potential reasons for women's reluctance to engage in novel work behavior. Most of the previous relevant literature suggests no gender difference, or a slight female advantage in creative skills and capacities (Baer & Kaufman, 2008; Kogan, 1974). Despite having similar creative abilities, previous work in a wide range of domains, such as psychology, and management, suggests that females' creative performance is inferior, compared to their male peers (Chavez-Eakle et al., 2006; Dul et al., 2011; Martín-Brufau & Corbalán, 2016).

Role congruity theory proposes that attributes traditionally associated with successful leaders, such as decisiveness and assertiveness, are incongruent with communal characteristics typically ascribed to women, such as compassion and care (Eagly & Karau, 2002). This misalignment between leadership and gender roles often leads to prejudice against female leaders. Novel work behavior, including the pursuit of novel research, is often perceived as a prototypically masculine activity due to its inherent riskiness, initiative-taking, and challenge-seeking nature, which are activities often associated with men. As a result, innovative work by females may face negative evaluations based on stereotypes. Drawing upon expectation states theory and gender status beliefs, Trapido (2022) identifies the "female penalty for novelty" phenomenon, which indicates that female authors' novel contributions receive less recognition than men's similar contributions. The author further provides empirical evidence that supports this hypothesis. Several empirical studies provide evidence of biases against women's novel contributions (Goldin & Rouse, 2000; Schmutz & Faupel, 2010). For instance, analyzing 1,503 popular music albums, Schmutz and Faupel (2010) find that there is a lower likelihood that female performers will receive cultural legitimacy, compared to their male counterparts, and their albums also have a more negligible probability of achieving consecrated status. Analyzing a sample of 895 ventures, Liao et al. (2023) find that investors are more likely to withhold funding support when women propose novel ventures, attributing this gender gap to gender role violations. Moreover, previous research proves insufficient institutional support for women's novel work. A recent study analyzing survey and employment data from 14,590 workers in the US reveals that women report less support for novelty in the workplace compared to men (Taylor et al., 2020).

These previous studies suggest that biases toward recognizing novel contributions by women and the lack of institutional support for their novel work may contribute to women's reluctance to generate novel work. This phenomenon is likely to be present in academia as well. However, empirical evidence supporting this prediction is currently lacking.

### 3.3 Heterogeneity of gender disparities in scientific novelty

The effect of gender on scientific novelty may coexist with that of other status differences. The magnitude of gender disparities in scientific novelty may vary based on the prestige of the university and the distribution of scientific novelty. Institutional resources and support are essential for scientists' research and may influence their research strategy choices (Allison & Long, 1990; Liu & Hu, 2022). Some researchers highlight the importance of gender differences concerning positions and resources, suggesting that the net difference between women and men in scientific outputs, such as productivity, might be negligible if there is no difference in positions and resources (Ceci & Williams, 2011; Council, 2010). If working at non-elite universities, such as teaching-intensive colleagues, resource scarcity affects female scientists more often than their male peers. Empirical evidence from Ceci and Williams (2011) suggests that male scientists produce 30% more publications than female scientists. However, the authors further find that, when comparing men and women who are tenured at R1 universities, the gender gap in publications decreases to 8%. Additionally, a recent report suggests that top-tier universities are creating work environments that are increasingly supportive and inclusive for female scientists. These institutions prioritize non-discrimination measures and actively promote gender equity, going beyond the basic requirements (Bothwell et al., 2022). Therefore, aligning with the view that highlights the importance of the prestige of a university in relation to gender disparities in science, and the potentially greater support provided by top-tier universities to female scientists, it is expected that gender disparities in research strategies, specifically pursuing novel research, may be mitigated among students in top-tier universities.

Gender disparities in scientific novelty may vary across different percentiles of novelty. Specifically, among doctoral theses with a lower level of novelty, gender disparities may be more pronounced compared to those among theses with a higher level of novelty. Previous research has shown that high achievers tend to have a stronger self-perception, higher self-concept, and greater self-confidence (Feather, 1989; McCoach & Siegle, 2001). Female students who produce highly novel work compared to their peers are considered high achievers in terms of generating new knowledge. These students likely possess unique skills, knowledge, ideas, and resources that enable them to create innovative knowledge compared to their peers who do not produce highly novel theses. On one hand, due to the merits and advantages that female high achievers have, they may exhibit higher self-confidence, which can lead to taking more risks (Krueger Jr & Dickson, 1994; Macko & Tyszka, 2009). Consequently, these female students may be less affected by the gender penalty associated with novelty and demonstrate a high level of scientific novelty in their theses. Therefore, the gender disparities in scientific novelty among doctoral theses might be more prominent among those with relatively lower levels of novelty but could diminish as the level of novelty increases toward the upper percentiles.

Previous studies indicate that differences in gender disparity exist regarding publications and citations. Particularly, among scientific elites, the underrepresentation of female scientists may be less significant, and they might even outperform their male counterparts (Chan & Torgler, 2020). This observation suggests that not all women are disadvantaged in science, and the level of achievement/status can influence the extent and direction of gender disparities in scientific output. We anticipate that this heterogeneity in gender disparities, considering the status of female scientists, also exists in terms of scientific novelty. Among high performers such as leaders, females are equally as, or even more competent and innovative than, males (Zenger & Folkman, 2019). This is attributed to the fact that female high performers often navigate additional obstacles in their careers compared to their male peers, developing skills such as resilience, adaptability, and creative thinking (Hampole et al., 2021). The investigation of high performers reveals a reversal in gender disparities. For example, a recent study on 29,809 management-track employees in a large retail chain revealed that females outperformed their male peers regarding performance ratings (Benson et al., 2021).

## 3.4 Research questions

The literature review mentioned above has identified several research gaps regarding the temporal development of scientific novelty, gender disparities in scientific novelty, and potential heterogeneities within those disparities. Firstly, there is a lack of understanding of how scientific novelty in doctoral theses has evolved over the past decades, as contrasting predictions based on the knowledge burden perspective and the promoting effect of emerging technology and societal needs make it unclear. Additionally, little is known about the research strategies pursued by early-career students that have led to novel research work over time. Therefore, the first research question proposed is:

RQ1: How has the scientific novelty of doctoral theses evolved over the past decades?

Despite extensive discussions and explorations on gender disparities in science, limited research focuses on their manifestations specifically in scientific novelty, with only a few exceptions (Trapido, 2022). Scientific novelty reflects one dimension of researchers' output, namely originality or novelty, and findings regarding gender disparities in productivity and other scientific activities cannot be directly applied to scientific novelty. Hence, the following research question is posed:

RQ2: Are there any gender disparities in the scientific novelty of doctoral theses?

To address RQ2, this study investigates students' gender and supervisors' gender and whether and how students' gender and supervisors' gender interact and influence the scientific novelty of doctoral theses.

As discussed in section 3.3, gender disparities in scientific novelty may vary depending on university prestige and the percentiles of scientific novelty in doctoral theses. The first type of heterogeneity arises from institutional differences in supporting female scientists, while the second type is attributed to higher confidence among female high achievers, leading to more risk-taking behaviors and an inclination to pursue highly novel research. However, empirical evidence supporting these hypotheses is lacking in the existing literature. As a result, the following research question is formulated:

RQ3: Do gender disparities in the scientific novelty of doctoral theses exhibit heterogeneities?

# 4 Data and Methods

## 4.1 Data source and processing

The major data source of this study is the Sciences and Engineering Collection of The ProQuest Dissertations & Theses Citation Index (hereafter PQDT). PQDT is the world's largest multidisciplinary dissertation database with records of over 5.5 million dissertations from thousands of universities worldwide. PQDT provides comprehensive historic and ongoing coverage for North American works since it is designated as an official offsite repository for the US Library of Congress. [2] The Science and Engineering collection of PQDT includes dissertations in a broad range of hard science disciplines, including the life sciences, mathematics, computer science, and engineering. We obtain 1,109,491 theses from the Science and Engineering collection of PQDT with publication years ranging from 1960 to 2016. PQDT offers information about dissertations including the author's name, advisor's name, university, subjects, year of publication, abstract, title, and so forth.

Each thesis in PQDT is assigned one subject or multiple subjects based on the author's selection from a list of subject categories provided by PQDT when they submit their work to the database. The subject(s) that are associated with each thesis reflect its disciplinary or topic attributes. The 1,109,491 theses we obtain from the Sciences and Engineering Collection of PQDT are found to be associated with 552 subjects that can be mapped to 22 broader disciplines based on PQDT's subject category classification scheme. The distribution of major scientific domains of theses in the original dataset is shown in Table S1. Because of sparse data points before 1980, we only investigate doctoral theses published from 1980 to 2016 to ensure data accuracy. We keep doctoral theses that are assigned to a subject or subjects in biology science, and health and medical science (hereafter biomedical sciences). The final dataset of this study contains 279,424 doctoral theses in biomedical sciences from 1980 to 2016. Table S2 provides basic statistics of the final dataset.

---
[2] https://about.proquest.com/en/products-services/pqdtglobal/

### 4.2 Predicting the gender information of students and advisors

The PQDT database lacks gender information for students and advisors. To infer their gender, we utilize the Gender-Guesser package, an open-source Python module. Gender-Guesser is a state-of-the-art tool for predicting gender based on first names. This tool has been widely employed in gender-related studies (Liu, Zhang, et al., 2022; Squazzoni et al., 2021; Zhang et al., 2022). Based on first names, the tool assigns prediction results into six categories: "female", "male", "mostly female", "mostly male", "andy", and "unknown". The categories of "mostly male" and "mostly female" indicate that a name is used by both genders but is more commonly associated with one. The "andy" category suggests a name used equally by males and females, while the "unknown" category refers to names not found in the gender dataset. The inferred gender information of students and advisors in the final dataset is shown in Table 1. We reclassify instances labeled as "mostly female" to "female" and those labeled as "mostly male" to "male". To investigate RQ2 and RQ3, our analysis will be limited to doctoral theses in which either the students or advisors have available predicted gender information.

**Table 1. The inferred gender information of students and advisors.**

|                   | Student   |                | Advisor   |                |
| ----------------- | --------- | -------------- | --------- | -------------- |
| Gender prediction | Frequency | Percentage (%) | Frequency | Percentage (%) |
| **Female**        | 98,597    | 45.1           | 50,998    | 22.09          |
| **Male**          | 103,585   | 47.38          | 141,008   | 61.08          |
| **Mostly female** | 11,122    | 5.09           | 5,954     | 2.58           |
| **Mostly male**   | 5,337     | 2.44           | 4,400     | 1.91           |
| **Andy**          | 11,881    | 4.25           | 3,290     | 1.43           |
| **Unknown**       | 48,902    | 17.5           | 25,221    | 10.92          |
| **Total**         | 279,424   | 100            | 230,871   | 100            |

### 4.3 Measuring scientific novelty of doctoral theses using a Bio-BERT model

Scientific novelty, or originality, newness, or atypicality, is the extent to which a scientific document contributes new theories, methods, data, or findings for subsequent studies. The concept of scientific novelty originated from Schumpeter's work on business cycles in the 1930s, where the recombinant nature of novelty was first outlined (Schumpeter, 1939). Nowadays, the perspective of recombinant scientific novelty has become the standard approach in studying innovation and has been applied by scholars across various disciplines to measure the novelty of scientific documents, including publications, patents, and grant proposals (Fleming, 2001; Simonton, 2003; Uzzi et al., 2013; Wang & Shibayama, 2022; Weitzman, 1998). As the volume of scientific data has rapidly increased, researchers have utilized text information or citation data to operationalize knowledge elements such as keywords (Boudreau et al., 2016; Chai & Menon, 2019), patent classes (Fleming, 2001), referenced articles or journals (Uzzi et al., 2013; Wang et al., 2017), and chemical entities (Foster et al., 2015). These knowledge elements are then used to quantify scientific novelty based on rare combinations of these elements. For example, Fleming (2001) suggests that patents combining previously uncombined technology classes can be considered novel technological creations. Boudreau et al. (2016) propose an indicator to assess the level of novelty in a research grant proposal by examining unique combinations of MeSH keywords. Uzzi et al. (2013) investigate the atypicality of pairs of referenced journals within a publication.

Drawing from the perspective of combinatorial novelty, Liu, Bu, et al. (2022) propose a methodology to assess scientific novelty of biomedical publications related to coronavirus. They utilize bio-entities as the fundamental knowledge element in their approach and employ a pre-trained Bio-BERT method to calculate the distance between these entities. The authors analyze all bio-entities within their dataset by pairing them up and measuring the distance between each pair of entities. To define novel entity combinations, the authors consider entity pairs whose distance falls within the upper $10^{th}$ percentile when compared to the distribution of distances among all entity pairs. They designate these unique entity pairs as novel combinations. The novelty score assigned to each publication is then determined by calculating the ratio of novel entity combinations to all possible entity combinations present within that publication.

In comparison to existing methods, the mentioned approach offers several advantages. This methodology considers the semantic relationship of knowledge components, leading to greater accuracy compared to superficially utilizing text information alone (Azoulay et al., 2011). Unlike methods that solely rely on cited articles/journals as knowledge units without delving deeper into their content (Uzzi et al., 2013; Wang & Shibayama, 2022), this study employs bio-entities as more detailed knowledge elements, enabling a more precise capture of knowledge combinations. Recent studies have utilized word-embedding models like word2vec to measure contextual and semantic distances between knowledge elements (Chiu & Baker, 2020; Yin et al., 2023). In this study, leveraging Bio-BERT, which incorporates domain-specific contexts and is well-suited for biomedical text-mining tasks, further enhances the method's effectiveness.

Considering the benefits above, this study adopts the approach proposed by Liu, Bu, et al. (2022) to assess the scientific novelty of biomedical doctoral theses through a five-step method.

**Extracting and disambiguating bio-entities**

We use BERN2 (Sung et al., 2022), an advanced neural biomedical tool, to extract biomedical entities from 279,424 doctoral theses. BERN2 consists of two main models: (1) Named Entity Recognition (NER), which identifies nine types of biomedical entities: gene/protein, disease, drug/chemical, species, mutation, cell line, cell type, DNA, and RNA, using a multi-task NER model; and (2) Named Entity Normalization (NEN), which links annotated entities to concept unique identifiers using rule-based and neural network-based NEN models. BERN2 outperforms existing biomedical text mining tools (Kim et al., 2019) by providing more efficient annotations. Using BERN2, we extract 1,519,599 annotated bio-entity names from the titles and abstracts of doctoral theses in the final dataset. These names are disambiguated and linked to 118,349 unique bio-entity IDs. The most frequently occurring bio-entity name associated with each ID in the biomedical doctoral theses is designated as the standard name. If there are multiple associated names with unequal occurrences, one is randomly chosen as the standard name. We then create pairings among the 118,349 unique bio-entity IDs based on their associations with the doctoral theses in the final dataset, resulting in 68,949,061 entity combinations.

**Measuring the distance of two bio-entities**

Using the standard names associated with the 118,349 unique bio-entity IDs obtained in the previous step, we convert each standard bio-entity name into a vector representation using a Bio-BERT model. We then calculate the distance between two bio-entities that are denoted by $i$ and $j$, $D_{i,j}$, for any entity combination that is generated from the doctoral theses using Equation 1.

$$D_{i,j} = 1 - CosSim_{i,j} \quad (1)$$

where $CosSim_{i,j}$ is the cosine similarity between entities $i$ and $j$ based on their corresponding vector representations that are obtained from the Bio-BERT model. The examples of an entity vector space for three theses based on the Bio-BERT model are shown in Figs. 1a-1b.

**Identifying novel entity pairs**

We develop a criterion to determine what qualifies as a novel combination of entities. To do this, we analyze the distribution of cosine distances among all pairs of entities in our dataset. If the cosine distance between the two constituent entities of a pair falls within the top 10% of this distribution, we consider it as a novel entity pairing. The 90th percentile of the distribution corresponds to a cosine distance of 0.279 (Fig. 1c). Any entity pair with a cosine distance greater than 0.279 is considered to be a novel combination. We further define **a novel thesis** as a doctoral thesis that includes at least one novel entity combination/pair.

**Calculating novelty scores**

To provide a nuanced evaluation of each doctoral thesis's scientific novelty, we introduce the **novelty score**. This score is calculated by determining the proportion of novel entity pairs out of the total number of possible entity pairs within a given thesis. The novelty score is bounded between 0 and 1, with a higher score indicating a greater degree of novelty. This metric provides a precise and continuous measure of the unique combinations of entities present in each thesis.

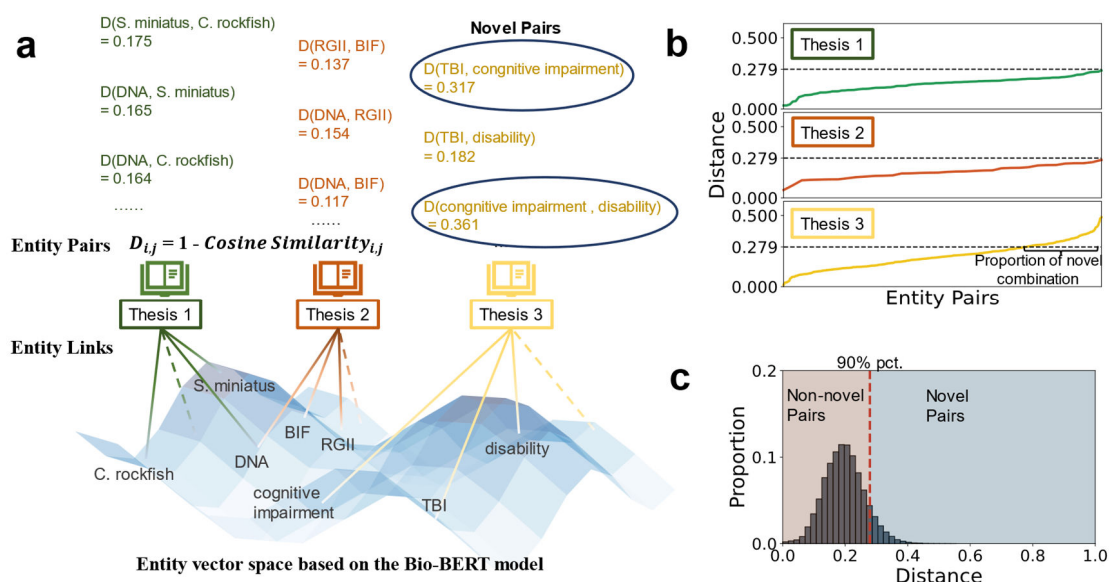

**Fig 1. The illustration of how to measure novelty scores for doctoral theses using the Bio-BERT model.** (**a**) An entity vector space containing all entities extracted from three sample doctoral theses based on Bio-BERT. (**b**) The distribution of cosine distances between entities for all entity pairs extracted from the three sample doctoral theses. Within each thesis, the entity pairs are ordered from left to right based on their cosine distance values. (**c**) The distribution of cosine distance for all entity pairs extracted from all doctoral theses in this study. If the cosine distance between the two constituent entities of an entity pair falls within the upper 10th percentile (i.e., 0.279) of this distribution, it is considered a novel entity pair.

### 4.4 Variables

To address RQs 2 and 3, the main independent variable is the gender of the student, which is denoted by $female\ student$. It equals 1 if the doctoral thesis is female-authored and 0 otherwise, based on the gender prediction of authors' names using Gender Guesser. The gender disparities in the scientific novelty of doctoral theses are not only reflected by students' gender but also manifest in advisors' gender. To reveal disparities more comprehensively, the gender of advisors is examined, using another explanatory variable, $female\ advisor$. To explore the possible interactive effect of the gender of students, and the gender of advisors, we generate an interaction term, $female\ student \times female\ advisor$.

Two dependent variables in this study are generated to measure the scientific novelty of doctoral theses. The first one is a binary variable and is used to measure if a doctoral thesis includes at least one novel bio-entity combination, based on our definition of a novel entity combination. We denote this variable by $novel\ combination$. It is 1 if a doctoral thesis contains at least one novel bio-entity combination, and 0 otherwise. The second dependent variable is a continuous variable that is denoted by $novelty\ sore$, which is applied to measure the fraction of novel entity combinations in each doctoral thesis.

We also include several control variables to obtain the net effect of gender on the scientific novelty of a doctoral thesis. Some characteristics of mentors, such as seniority, and the mentoring experience, may be related to the mentoring outcome (Wuestman et al., 2023; Xing et al., 2022), and influence the scientific novelty of a thesis. We use *career age* to proxy for mentors' seniority, which is the number of years that have elapsed since the student's mentor supervised his/her first mentee in PQDT. We include *mentee number*, which is defined as the accumulative number of mentees the mentor has supervised in PQDT. The length of a scientific document is considered relevant to scientific novelty (Liang et al., 2023) and we take the length of each thesis (*thesis length*) into account. Interdisciplinary research is believed to foster scientific novelty (D'este et al., 2019; Fontana et al., 2020), and we use whether or not the thesis is assigned multiple subjects in PQDT (*interdisciplinary*) to measure the interdisciplinarity of each thesis. Organizational context is an important influential factor in the quality of students' theses, and organizational support plays a key role in nurturing students' novel research (Wang & Shibayama, 2022). Based on the Carnegie Classification of Institutions of Higher Education,[3] we categorize US universities in the final dataset into three categories: R1 university (university with very high research intensity), R2 university (university with high research intensity), and other types of university. A categorical variable that refers to this classification is used as a control (*university prestige*) in this study. This variable is coded on a scale of 1 to 3, with a university prestige of 1 denoting an R1 university, a university prestige of 2 representing an R2 university, and a university prestige of 3 indicating other types of universities. Summary statistics of variables and the correlation coefficient matrix are shown in Tables 1 and 2, respectively.

**Table 2. Summary statistics of variables.**

| Variable | Obs. | Mean | Std. dev. | Min | Max |
|---|---|---|---|---|---|
| **Novel combination** | 218,641 | 0.59 | 0.492 | 0 | 1 |

---
[3] https://carnegieclassifications.acenet.edu/

|                     |         |       |       |   |       |
|---------------------|---------|-------|-------|---|-------|
| **Novelty score**   | 218,641 | 0.083 | 0.121 | 0 | 1     |
| **Female student**  | 218,641 | 0.502 | 0.5   | 0 | 1     |
| **Female advisor**  | 158,838 | 0.293 | 0.455 | 0 | 1     |
| **Career age**      | 177,819 | 0.792 | 1.023 | 0 | 3.555 |
| **Mentee number**   | 177,819 | 0.588 | 0.719 | 0 | 3.738 |
| **Thesis length**   | 216,328 | 5.024 | 0.63  | 0 | 9.044 |
| **Interdisciplinary** | 218,641 | 0.486 | 0.5 | 0 | 1     |

Notes: all the continuous variables, except novelty score, are naturally log-transformed.

**Table 3. Spearman's rank correlation coefficient matrix of variables.**

|   |                    | (1)      | (2)      | (3)     | (4)      | (5)      | (6)      | (7)     |
|---|--------------------|----------|----------|---------|----------|----------|----------|---------|
| (1) | **Novel combination** | 1      |          |         |          |          |          |         |
| (2) | **Novelty score**    | 0.883**  | 1        |         |          |          |          |         |
| (3) | **Female student**   | -0.084** | -0.071** | 1       |          |          |          |         |
| (4) | **Female advisor**   | -0.115** | -0.090** | 0.196** | 1        |          |          |         |
| (5) | **Career age**       | 0.077**  | 0.054**  | 0.007** | -0.030** | 1        |          |         |
| (6) | **Mentee number**    | 0.077**  | 0.054**  | 0.013** | -0.013** | 0.901**  | 1        |         |
| (7) | **Thesis length**    | 0.010**  | 0.002    | 0.004   | 0.015**  | 0.001    | -0.008** | 1       |
| (8) | **Interdisciplinary**| -0.189** | -0.141** | 0.046** | 0.001    | -0.083** | -0.103** | 0.033** |

Notes: all the continuous variables, except novelty score, are naturally log-transformed. * $p<0.05$; ** $p<0.01$.

## 4.5 Regression analyses

To address RQ1, we mainly apply descriptive analyses, t-test analyses, and simple linear regression analyses. We use logistic regression models, fractional logistic regression models, subgroup regression analyses, and quantile regression analyses to address RQ2 and RQ3.

## 4.5.1 Logistic and fractional logistic regression models

To address RQ2, the two dependent variables are estimated using Equation 2.

$$Dependent\ variable_t = \alpha + \beta_1 female\_student_t + \beta_2 female\_advisor_t + \beta_3 female\_student_t \times female\_advisor_t + Controls + Y_t + \varepsilon\ (2)$$

when the dependent variable is *novel combination*, we use a standard logistic regression model to estimate Equation 2 because it is a binary variable. A fractional logit regression model is applied when the dependent variable is *novelty score*; that is, a continuous variable restricted within the bounded range of 0 to 1. *Controls* include a few control variables that may be related to scientific novelty of students' theses, which is introduced in section 4.4. $Y_t$ refers to year-fixed effects, which are used to control time-variant unobserved changes, such as policy changes that support scientific novelty. The mean-variance inflation factor (VIF) obtained is 2.03, which is significantly lower than the threshold value of 5. This indicates that there are no issues of multicollinearity present in the regression model.

To investigate the first aspect of RQ3, about the heterogeneity of gender disparities in scientific novelty based on university prestige, we employ subgroup regression analyses. More specifically, we conduct separate analyses for R1 and non-R1 universities, comparing the magnitude of gender disparities in scientific novelty within each group.

### 4.5.2 Quantile regression models

To address the second part of RQ3, i.e., the heterogeneous gender disparities in scientific novelty of doctoral theses concerning the different percentiles of scientific novelty scores, we employ quantile regression models. This approach offers a more comprehensive statistical analysis opportunity compared to the traditional mean regression model (Koenker & Bassett Jr, 1978; Koenker & Hallock, 2001). Quantile regression allows estimation of the relationship between explanatory variables and the conditional quantity of the dependent variable without assuming a specific conditional distribution. By accounting for potential unobserved heterogeneity, this approach enables investigation into different aspects of the dependent variable's distribution. Including quantile regression models in the analysis of gender disparity in the scientific novelty of doctoral theses allows exploration of the heterogeneous gender differences across various quantiles (e.g., ranging from 0.1 to 0.9) of the scientific novelty distribution.

We utilize quantile regression models to estimate different quantile points within the distribution of scientific novelty scores in doctoral theses. Because the $40^{th}$ percentile of the distribution of scientific novelty scores is 0, there will be no gender differences in scientific novelty scores. We choose to fit a multivariate quantile regression on the $50^{th}$, $60^{th}$, $70^{th}$, $80^{th}$, and $90^{th}$ percentiles of the distribution of scientific novelty scores. We consider the 0.5-, 0.6-, 0.7-, 0.8- and 0.9-quantile points. The linear quantile regression model is defined by Equation 3.

$$Q_{y_t}(\tau) = \sum_{i=1}^{k} \beta_{\tau,i} x_{ti} \quad (3)$$

where $\beta_{\tau,i}$ is an unknown parameter and $Q_{y_t}(\tau)$ denotes the $\tau^{th}$ conditional quantile of scientific novelty scores that is denoted by $y_t$ ($0 < \tau < 1$), i.e., the $\tau^{th}$ quantile of $y_t$ given $x_t$. $Q_{y_t}(0.5)$ refers to the distribution mean. The regression coefficients are estimated using an asymmetric absolute loss function.

## 5 Results

### 5.1 The downward trend of scientific novelty in doctoral theses over the past four decades

There has been a decreasing trend in the scientific novelty of doctoral theses in the biomedical field over the past four decades. A temporal pattern reveals a noticeable decline in the fraction of novel theses and the average novelty score of doctoral theses as time progresses. This trend is evidenced by the Empirical Cumulative Distribution Function (ECDF) curves, with the theses from earlier periods located on the right or bottom side (Figs. 2a and 2e,). Welch's t-tests reveal significant increases in both the fraction of novel theses and the average scientific novelty scores of doctoral theses in recent decades. Notably, there have been significant linear decreases in these two variables since 1980 ($p < 0.001$, coefficient of year=-0.001, Fig. 2c; $p < 0.001$, coefficient of year=-0.004, Fig. 2g). These downward trends are consistent across universities of varying prestige (Figs. 2d and 2h) and across genders (Figs. 2b and 2c).

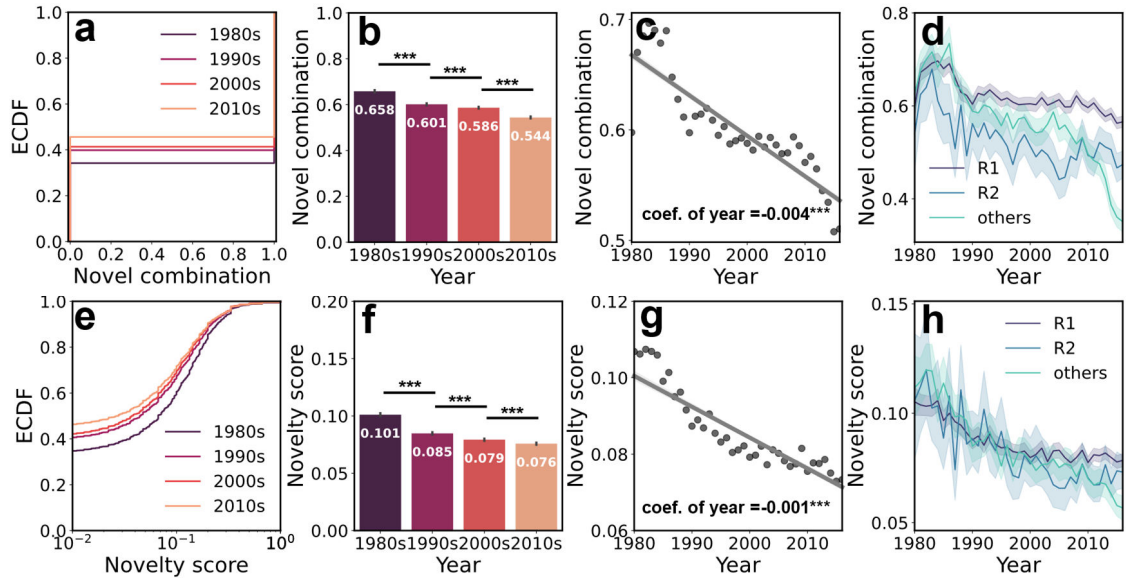

**Fig 2. The temporal evolution of scientific novelty in doctoral theses.** (**a**, **e**) ECDF of novel combination (**a**) and novelty score (**e**) in each period. (**b**, **f**) The fraction of novel theses (**b**) and the mean of novelty scores (**f**) in each period. Two-tailed Welch's t-tests are employed to compare novel combination and novelty score in two consecutive periods. (**c**, **g**) Linear regressions predicting the fraction of novel theses (**c**) from year and the average novelty score (**g**) from year. (**d**, **h**) The fraction of novel theses and the average novelty score over the years across universities of different prestige levels. *** $p < 0.001$. The error bars/shaded areas indicate the 95% confidence intervals.

## 5.2 Gender disparities in producing novel doctoral theses

We identify a persistent gender disparity in the scientific novelty of doctoral theses. Specifically, female students are less likely to produce a novel doctoral thesis. Additionally, the average novelty score of female-authored doctoral theses is significantly lower compared to that of male-authored ones.

Our findings reveal that 63.14% of male students produced a novel doctoral thesis, a significantly higher percentage than that of female students by 8.27% ($p < 0.001$, Welch's t-test, Fig. 3b). The distribution of novelty scores for doctoral theses by female and male students (Fig. 3a) also indicates that a larger proportion of female-authored theses received a score of 0 or a low value in terms of scientific novelty compared to male-authored theses. Fig. 3c illustrates that doctoral theses authored by female students demonstrate a lower level of novelty when compared to those authored by male students, as the average novelty score of female-authored theses is 0.0086 lower ($p < 0.001$). This accounts for approximately 10.33% of the mean novelty score of all doctoral theses. The gender disparity in scientific novelty persists across various periods. Figs. 3b and 3c depict a consistent gender gap in both the percentage of novel doctoral theses produced by female and male students and the average novelty scores of female-authored and male-authored theses, spanning the past four decades.

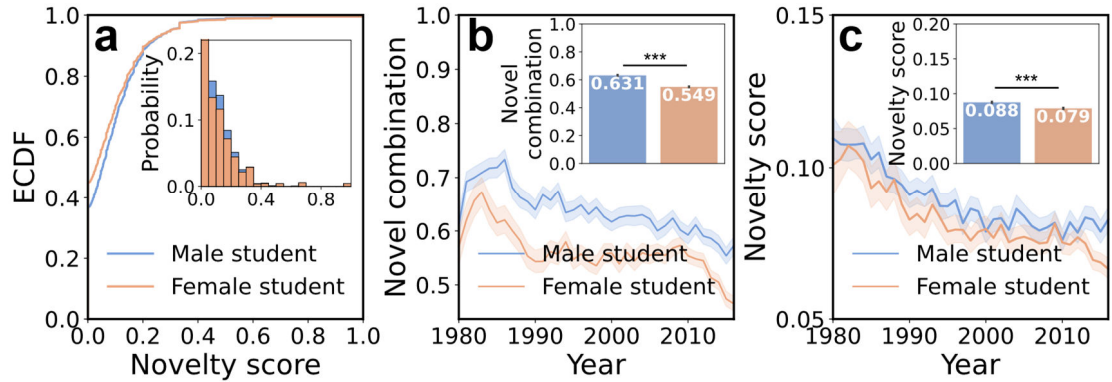

**Fig 3. Gender disparities in the scientific novelty of doctoral theses based on students' gender. (a)** ECDF and distributions of novelty scores for male and female students. **(b)** The fraction of novel theses for female and male students over the years. A two-tailed Welch's t-test is employed to compare female and male students' probability of producing a novel thesis. **(c)** The average scientific novelty of doctoral theses for female and male students over the years. A two-tailed Welch's t-test is employed to compare the average scientific novelty of female-authored and male-authored doctoral theses. *** $p < 0.001$. The error bars/shaded areas indicate the 95% confidence intervals.

The aforementioned gender disparity also exists when examining the gender of advisors. Fig. 4a suggests that a larger proportion of students who are supervised by female advisors tend to produce theses that do not contain any novel combinations. Among the students under the supervision of male advisors, more than 60% produced a novel doctoral thesis, which is far higher than the figure for female advisors (i.e., 48.35%) ($p < 0.01$, Welch's t-test, Fig. 4b). This finding indicates that, under the supervision of male advisors, students may be more likely to produce a novel doctoral thesis, compared to if they are supervised by female advisors. Additionally, the average novelty score of doctoral theses authored by students under the supervision of male advisors is significantly higher than that for female advisors ($p < 0.01$, Welch's t-test, Fig. 4c). The prominent gender differences concerning the proportion of novel doctoral theses and the average novelty score of doctoral theses from the perspective of advisors' gender were persistent across different periods (Figs. 4b and 4c).

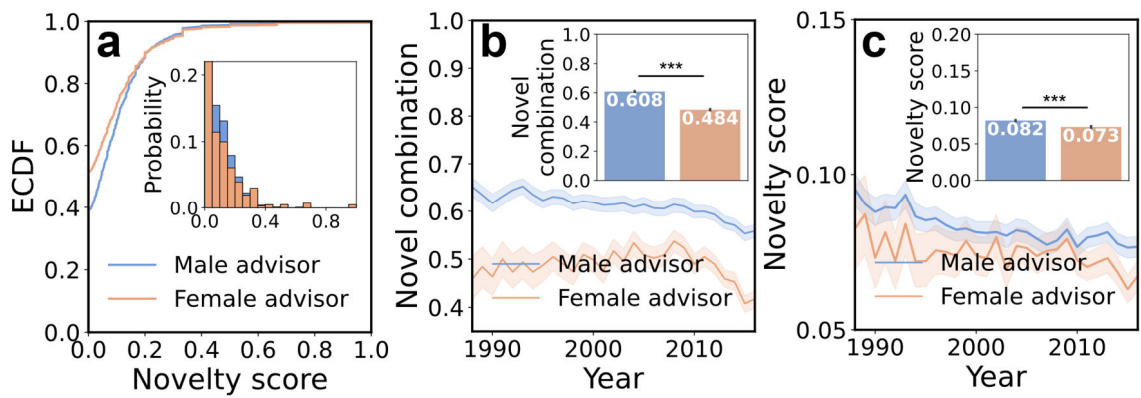

**Fig 4. Gender disparities in scientific novelty among doctoral theses based on advisors' gender. (a)** ECDF and distributions of novelty scores for male and female advisors. **(b)** The fraction of novel theses for female and male advisors over the years. A two-tailed Welch's t-test is employed to compare the probability of students who are supervised by female advisors or male advisors producing a novel thesis. **(c)** The average scientific novelty of doctoral theses for students who are supervised by female or male advisors over the years. A two-tailed Welch's t-test is employed to compare the average scientific novelty of doctoral theses produced by students who are supervised by female and male advisors. *** $p < 0.001$. The error bars/shaded areas indicate the 95% confidence intervals.

The observed gender discrepancies regarding the probability of producing a novel doctoral thesis still hold when performing regression analyses with a variety of influential factors controlled. Compared to their male peers, female students are less likely to produce a novel doctoral thesis by 7.0% (marginal effect, $p < 0.01$, coefficient=−0.290, odds ratio=0.748, columns 1 and 2 in Table 4, Model 1 of Fig. 5, Fig. 6a) when control variables, such as career age, mentoring experience of their advisors, and the characteristics of doctoral theses, such as thesis length, whether or not the doctoral thesis is interdisciplinary, and the research intensity of universities, are set to their means. Additionally, under the supervision of female advisors, the probability of students producing a doctoral thesis decreases by 10.4 % (marginal effect, $p < 0.01$, coefficient=−0.424, odds ratio=0.654, columns 3 and 4 in Table 4, Model 2 of Fig. 5, Fig. 6b) when all the control variables are set to their means.

We find a significant interaction effect of students' gender and advisors' gender on the probability of students producing a novel thesis. The coefficient of the interaction term of students' gender and advisors' gender on whether or not the student produces a novel thesis is significantly negative (coefficient: -0.460, odds ratio: 0.631, $p < 0.01$, columns 5 and 6 in Table 4, Model 3 of Fig. 5). From Fig. 6c, when being supervised by male advisors, the gender gap in the possibility of producing a novel thesis between female students and male students is slight, and only approaches 2%. However, the gender gap significantly widens and increases to around 13% if students are supervised by female advisors. Those findings suggest that being supervised by female advisors may strengthen the gender disparity in the probability of producing a novel thesis.

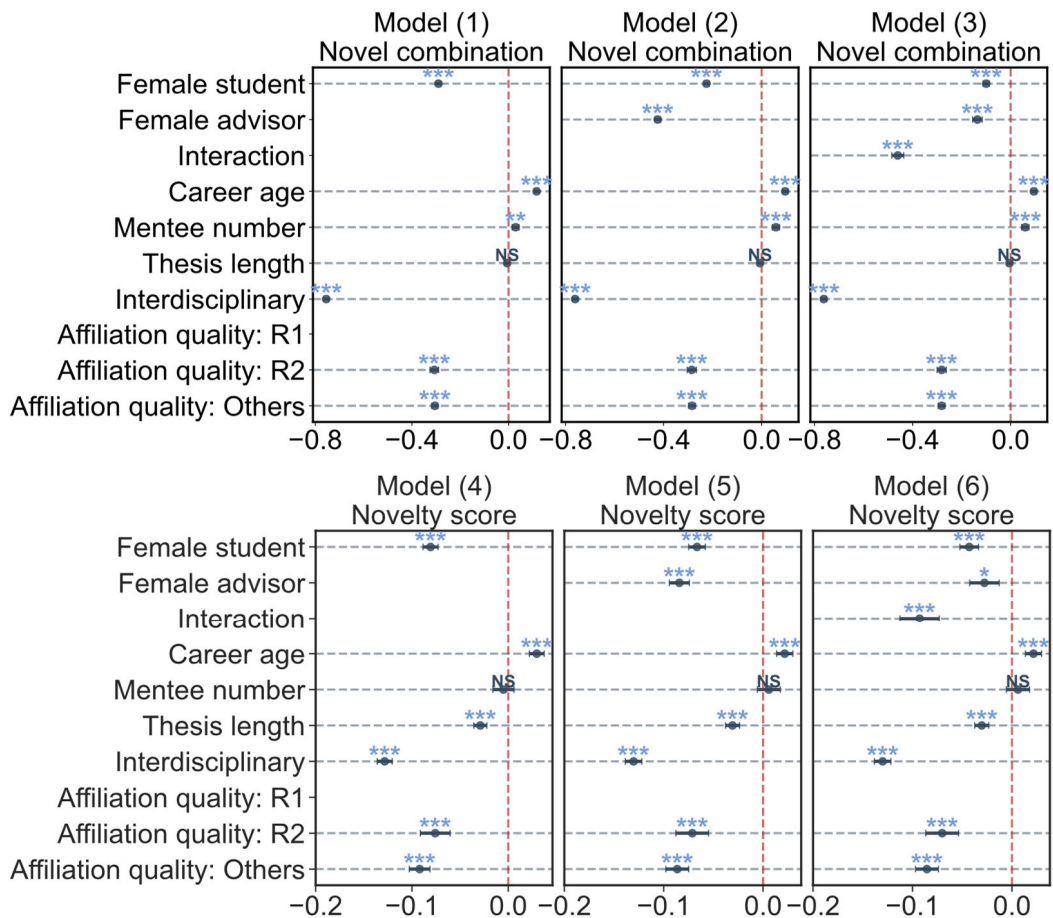

**Fig. 5. The estimated regression coefficients of variables from models 1 to 6 in Table 4.** Interaction indicates the interaction term between female student and female advisor. *** p<0.01, ** p<0.05, * p<0.1. NS refers to not significant. The error bars represent the upper and lower bounds of 95% confidence

intervals.

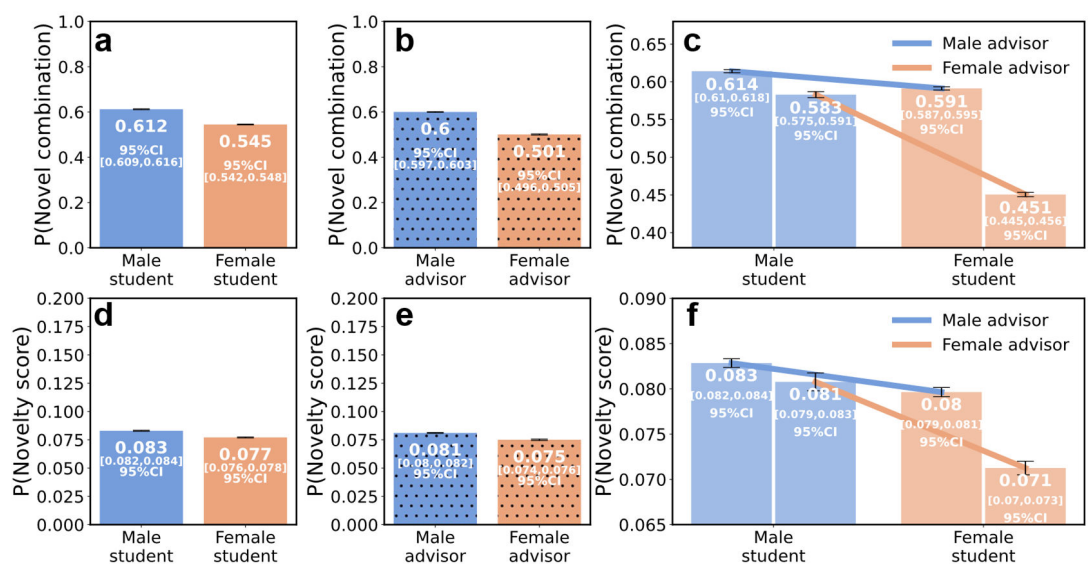

**Fig 6. The linear prediction of the probability of students producing a novel thesis and novelty scores of doctoral theses based on students' gender, advisors' gender, and their interaction term.** (**a**, **b**, **c**) The predicted dependent variable, i.e., the probability of students producing a novel thesis, when all covariates are set to their means. (**d**, **e**, **f**) The predicted dependent variable, i.e., scientific novelty scores of doctoral theses, when all covariates are set to their means. The blue line indicates being under the supervision of male advisors, and the orange line indicates being under the supervision of female advisors (**c**, **f**). The error bars represent the upper and lower bounds of 95% confidence intervals.

The disadvantage of female students is also found if we investigate the scientific novelty score of students' doctoral theses by applying fractional logistic regression models. The regression result suggests that, on average, doctoral theses authored by female students have a significantly smaller scientific novelty score (coefficient= -0.081, $p < 0.01$; odds ratio=0.923, p<0.01, columns 1 and 2 of Table 5, Model 4 of Fig. 5, Fig. 6d) than those authored by male students by -0.0059 (marginal effect, p<0.01), that is approximately 7% of the mean of scientific novelty for all doctoral theses in the final dataset of this study. Furthermore, students under supervision by female advisors produce doctoral theses with a significantly smaller scientific novelty by 0.084 (coefficient=-0.084, $p < 0.01$; odds ratio=0.919, $p < 0.01$, columns 3 and 4 of Table 5, Model 5 of Fig. 5, Fig. 6e), relative to those that are authored by students who are supervised by male advisors. The significant interaction effect between female students and female advisors (coefficient=-0.093, $p < 0.01$; odds ratio=0.911, $p < 0.01$, columns 5 and 6 of Table 5, Model 6 of Fig. 5, Fig. 6f) on novelty scores of doctoral theses indicates that the supervision by female advisors may intensify the gender disparity in scientific novelty scores of students' doctoral theses.

**Table 4. The estimated relationship between students' gender and the probability of students producing a novel doctoral thesis (novel combination) using logistic regression models.**

| Variable | (1) | (2) | (3) | (4) | (5) | (6) |
|---|---|---|---|---|---|---|
| | Model 1 | | Model 2 | | Model 3 | |
| | Coefficient | Odds ratio | Coefficient | Odds ratio | Coefficient | Odds ratio |
| **Female student** | -0.290*** | 0.748*** | -0.225*** | 0.799*** | -0.100*** | 0.905*** |
| | (0.010) | (0.008) | (0.011) | (0.009) | (0.013) | (0.012) |
| **Female advisor** | | | -0.424*** | 0.654*** | -0.135*** | 0.873*** |
| | | | (0.012) | (0.008) | (0.019) | (0.017) |

| | | | | | -0.460*** | 0.631*** |
|---|---|---|---|---|---|---|
| **Female student × Female advisor** | | | | | (0.024) | (0.015) |
| **Career age** | 0.116*** | 1.123*** | 0.096*** | 1.101*** | 0.095*** | 1.100*** |
| | (0.010) | (0.011) | (0.011) | (0.012) | (0.011) | (0.012) |
| **Mentee number** | 0.029** | 1.030** | 0.058*** | 1.060*** | 0.060*** | 1.062*** |
| | (0.014) | (0.015) | (0.015) | (0.016) | (0.015) | (0.016) |
| **Thesis length** | -0.006 | 0.994 | -0.007 | 0.993 | -0.005 | 0.995 |
| | (0.009) | (0.009) | (0.010) | (0.010) | (0.010) | (0.010) |
| **Interdisciplinary** | -0.755*** | 0.470*** | -0.761*** | 0.467*** | -0.762*** | 0.467*** |
| | (0.010) | (0.005) | (0.011) | (0.005) | (0.011) | (0.005) |
| **R1 university (baseline)** | - | - | - | - | - | - |
| **R2 university** | -0.307*** | 0.736*** | -0.285*** | 0.752*** | -0.281*** | 0.755*** |
| | (0.018) | (0.013) | (0.019) | (0.014) | (0.019) | (0.014) |
| **Other types** | -0.305*** | 0.737*** | -0.284*** | 0.753*** | -0.281*** | 0.755*** |
| | (0.013) | (0.010) | (0.014) | (0.010) | (0.014) | (0.010) |
| **Year Fixed Effect** | Yes | Yes | Yes | Yes | Yes | Yes |
| **Observations** | 173,534 | 173,534 | 155,148 | 155,148 | 155,148 | 155,148 |
| **pseudo $R^2$** | 0.0389 | 0.0389 | 0.0459 | 0.0459 | 0.0475 | 0.0475 |

**Notes:** Robust seeform are in parentheses. *** p<0.01, ** p<0.05, * p<0.1.

**Table 5. The estimated relationship between students' gender and novelty scores of students' doctoral thesis (novel score) using fractional logistic regression models.**

| | (1) | (2) | (3) | (4) | (5) | (6) |
|---|---|---|---|---|---|---|
| **Variable** | Model 1 | | Model 2 | | Model 3 | |
| | Coefficient | Odds ratio | Coefficient | Odds ratio | Coefficient | Odds ratio |
| **Female student** | -0.081*** | 0.923*** | -0.066*** | 0.936*** | -0.043*** | 0.958*** |
| | (0.008) | (0.007) | (0.008) | (0.008) | (0.009) | (0.009) |
| **Female advisor** | | | -0.084*** | 0.919*** | -0.028* | 0.973* |
| | | | (0.010) | (0.009) | (0.015) | (0.014) |
| **Female student × Female advisor** | | | | | -0.093*** | 0.911*** |
| | | | | | (0.020) | (0.018) |
| **Career age** | 0.030*** | 1.030*** | 0.022*** | 1.022*** | 0.022*** | 1.022*** |
| | (0.008) | (0.008) | (0.008) | (0.008) | (0.008) | (0.008) |
| **Mentee number** | -0.005 | 0.995 | 0.006 | 1.006 | 0.006 | 1.006 |
| | (0.011) | (0.011) | (0.012) | (0.012) | (0.012) | (0.012) |
| **Thesis length** | -0.029*** | 0.971*** | -0.031*** | 0.970*** | -0.030*** | 0.970*** |
| | (0.007) | (0.007) | (0.007) | (0.007) | (0.007) | (0.007) |
| **Interdisciplinary** | -0.128*** | 0.880*** | -0.130*** | 0.878*** | -0.130*** | 0.878*** |
| | (0.008) | (0.007) | (0.008) | (0.007) | (0.008) | (0.007) |
| **R1 university (baseline)** | - | - | - | - | - | - |
| **R2 university** | -0.076*** | 0.927*** | -0.071*** | 0.931*** | -0.070*** | 0.932*** |
| | (0.015) | (0.014) | (0.016) | (0.015) | (0.016) | (0.015) |
| **Other types** | -0.092*** | 0.912*** | -0.086*** | 0.917*** | -0.085*** | 0.918*** |
| | (0.011) | (0.010) | (0.012) | (0.011) | (0.012) | (0.011) |
| **Year Fixed Effect** | Yes | Yes | Yes | Yes | Yes | Yes |
| **Observations** | 173,535 | 173,535 | 155,149 | 155,149 | 155,149 | 155,149 |
| **pseudo $R^2$** | 0.00174 | 0.00174 | 0.00196 | 0.00196 | 0.00201 | 0.00201 |

**Notes:** Robust seeform are in parentheses. *** p<0.01, ** p<0.05, * p<0.1.

### 5.3 The heterogeneities of gender disparities in the scientific novelty of doctoral

theses

### 5.3.1 Prestige of university

We observe further heterogeneities in the gender disparity of scientific novelty in doctoral theses, particularly concerning university prestige. Subgroup regression analyses reveal that, in R1 universities, compared to male peers, female students have a lower probability of producing a novel thesis by -0.034, (marginal effect, coefficient=-0.155, odds ratio=0.856, $p < 0.01$, columns 1 to 2 of Table 6). This represents 6.27% of the mean of the novel combination. On the other hand, in non-R1 universities, when controlling for other variables set at their means, the marginal effect of female students on the probability of producing a novel thesis is -0.010 (marginal effect, coefficient=-0.400, odds ratio=0.670, $p < 0.01$, columns 3 to 4 of Table 6), which accounts for 17.9% of the mean of the novel combination. These findings indicate a higher degree of gender differences in the production of novel doctoral theses in universities with a lower level of prestige (-6.27% vs -17.9%).

When analyzing the scientific novelty scores of doctoral theses, we find an amplified gender disparity in non-R1 universities as well. In R1 universities, female-authored theses have lower scientific scores than male-authored ones by -0.003 (marginal effect, coefficient=-0.046, odds ratio=0.955, $p < 0.01$, columns 5 to 6 of Table 6), equivalent to 4.1% of the mean scientific novelty scores across all the theses analyzed in this study. This disparity becomes more pronounced at -0.008 (marginal effect, coefficient=-0.119, odds ratio=0.888, $p < 0.01$, columns 7 to 8 of Table 6), approximately 10% of the mean scientific novelty scores for all theses in the final dataset.

**Table 6. The estimated relationship between students' gender and novelty scores of students' doctoral theses (novel score) using fractional logistic regression models.**

|  | (1) | (2) | (3) | (4) | (5) | (6) | (7) | (8) |
|---|---|---|---|---|---|---|---|---|
|  | \multicolumn{4}{c}{Novel combination} | \multicolumn{4}{c}{Novelty score} |
| Variable | Model1 | | Model2 | | Model3 | | Model4 | |
|  | R1 university | | Non-R1 university | | R1 university | | Non-R1 university | |
|  | Coefficient | Odds ratio | Coefficient | Odds ratio | Coefficient | Odds ratio | Coefficient | Odds ratio |
| **Female student** | -0.155*** | 0.856*** | -0.400*** | 0.670*** | -0.046*** | 0.955*** | -0.119*** | 0.888*** |
|  | (0.013) | (0.011) | (0.021) | (0.014) | (0.009) | (0.009) | (0.017) | (0.015) |
| **Female advisor** | -0.334*** | 0.716*** | -0.602*** | 0.548*** | -0.051*** | 0.951*** | -0.161*** | 0.852*** |
|  | (0.014) | (0.010) | (0.022) | (0.012) | (0.011) | (0.011) | (0.020) | (0.017) |
| **Career age** | -0.025* | 0.976* | 0.303*** | 1.353*** | -0.008 | 0.992 | 0.080*** | 1.083*** |
|  | (0.013) | (0.012) | (0.020) | (0.027) | (0.009) | (0.009) | (0.017) | (0.018) |
| **Mentee number** | 0.278*** | 1.320*** | -0.314*** | 0.730*** | 0.044*** | 1.045*** | -0.049** | 0.952** |
|  | (0.018) | (0.024) | (0.027) | (0.020) | (0.013) | (0.013) | (0.025) | (0.024) |
| **Thesis length** | -0.004 | 0.996 | -0.020 | 0.980 | -0.027*** | 0.974*** | -0.037*** | 0.964*** |
|  | (0.012) | (0.012) | (0.017) | (0.017) | (0.009) | (0.009) | (0.011) | (0.011) |
| **Interdisciplinary** | -0.683*** | 0.505*** | -0.939*** | 0.391*** | -0.097*** | 0.907*** | -0.204*** | 0.816*** |
|  | (0.013) | (0.006) | (0.021) | (0.008) | (0.009) | (0.009) | (0.017) | (0.014) |
| **Year FE** | YES | YES | YES | YES | YES | YES | YES | YES |
| **Observations** | 113,716 | 113,716 | 43,337 | 43,337 | 113,720 | 113,720 | 43,338 | 43,338 |
| **pseudo $R^2$** | 0.034 | 0.034 | 0.075 | 0.075 | 0.001 | 0.001 | 0.006 | 0.006 |

**Notes**: Robust seeform are in parentheses. *** p<0.01, ** p<0.05, * p<0.1.

## 5.3.2 The distribution of scientific novelty

The findings from quantile regression analyses reveal that gender differences in the scientific novelty of doctoral theses vary across different quantiles of the distribution of scientific novelty scores. Specifically, gender disparities are less pronounced at higher percentiles of the distribution (Fig.7). Examining the median (50th percentile) through multivariate quantile regressions, the coefficient for *female student* is -0.009 ($p < 0.01$, column 1 of Table 7), indicating that female-authored theses have a lower scientific novelty score compared to male-authored ones by approximately -0.009, equivalent to 10.8% of the mean scientific novelty score for all theses. The regression coefficient for *female student* on the 60th percentile is -0.013 ($p < 0.01$, column 2 of Table 7), illustrating a gender difference of -0.013 in scientific novelty scores at this specific percentile. Subsequently, the regression coefficients decrease to -0.010 ($p < 0.01$), -0.006 ($p < 0.01$), and -0.003 ($p < 0.01$) for the 70th, 80th, and 90th percentiles of the scientific novelty distribution (columns 3 to 5 of Table 7), respectively. These results suggest that gender differences reach their lowest point at high percentiles of the distribution, indicating a reduced gender disparity in scientific novelty on the upper end of the scale.

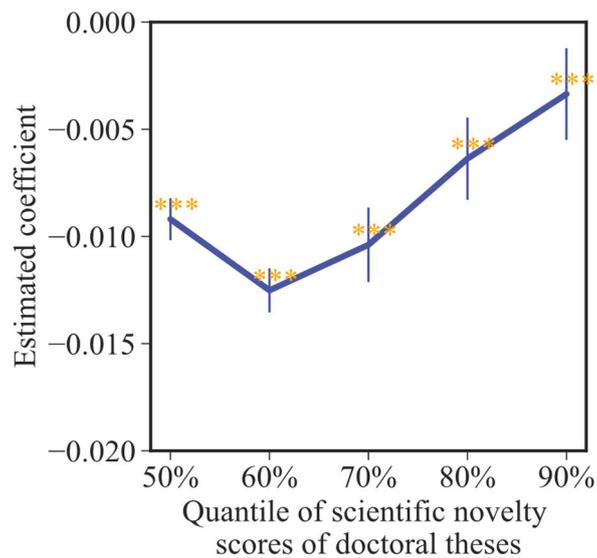

**Fig 7. The quantile estimation of the relationship between whether or not the student is female and scientific novelty scores of doctoral theses.** The y-axis indicates the estimated coefficients and the x-axis indicates the quantiles of the distribution of scientific novelty scores. *** p<0.01, ** p<0.05, * p<0.1. The error bars represent the upper and lower bounds of 95% confidence intervals.

**Table 7. Quantile regression results about the relationship between whether or not the student is female and the scientific novelty scores of their doctoral theses.**

|  | (1) Q50 | (2) Q60 | (3) Q70 | (4) Q80 | (5) Q90 |
| --- | --- | --- | --- | --- | --- |
| Female student | -0.009*** | -0.013*** | -0.010*** | -0.006*** | -0.003*** |
|  | (0.000) | (0.001) | (0.001) | (0.001) | (0.001) |
| Female advisor | -0.013*** | -0.025*** | -0.017*** | -0.009*** | 0.002 |
|  | (0.001) | (0.001) | (0.001) | (0.001) | (0.001) |
| Career age | 0.003*** | 0.006*** | 0.005*** | 0.004*** | 0.000 |
|  | (0.000) | (0.000) | (0.001) | (0.001) | (0.001) |
| Mentee number | 0.002*** | -0.000 | -0.002 | -0.002 | -0.001 |
|  | (0.000) | (0.000) | (0.001) | (0.001) | (0.001) |
| Thesis length | 0.000 | 0.000 | -0.002** | -0.003*** | -0.007*** |
| | (0.000) | (0.000) | (0.001) | (0.001) | (0.001) |
| Interdisciplinary | -0.034*** | -0.031*** | -0.020*** | -0.009*** | 0.002* |
| | (0.000) | (0.001) | (0.001) | (0.001) | (0.001) |
| R1 university (baseline) | - | - | - | - | - |
| R2 university | -0.006*** | -0.016*** | -0.013*** | -0.008*** | 0.000 |
| | (0.000) | (0.001) | (0.002) | (0.002) | (0.002) |
| Other types | -0.005*** | -0.016*** | -0.013*** | -0.009*** | -0.003*** |
| | (0.000) | (0.001) | (0.001) | (0.001) | (0.001) |
| Year Fixed Effect | Yes | Yes | Yes | Yes | Yes |
| Observations | 155,149 | 155,149 | 155,149 | 155,149 | 155,149 |
| pseudo $R^2$ | 0.0329 | 0.0220 | 0.0101 | 0.00399 | 0.00255 |

**Notes**: Robust seeform are in parentheses. *** $p<0.01$, ** $p<0.05$, * $p<0.1$.

### 5.4 Robustness checks

We conducted several robustness checks, and the overall findings remain largely unchanged. Firstly, to ensure accurate gender prediction, we only considered observations with clear predicted gender information (i.e., "female" or "male") and excluded those labeled as "mostly female" or "mostly male". We then reanalyze the data, and the regression results are presented in Tables S4 to S6, showing consistent major results.

Secondly, to mitigate any potential bias from the gender prediction tool, we employed an alternative widely used technique called Genderize.io (Santamaría & Mihaljević, 2018; Sebo, 2021) to infer the gender information of students and advisors. We retained only those observations with a predicted gender probability of 0.9 and above from the Genderize.io database for improved prediction accuracy. The predicted gender information for students and authors is displayed in Table S3. Once again, we obtained consistent major results, as demonstrated in Tables S7 to S9.

Lastly, to further validate our method of assessing scientific novelty and ensure it does not impact the main findings, we utilized another measurement proposed by Azoulay et al. (2011) to quantify scientific novelty of doctoral theses in biomedical sciences. This method calculates the average age of MeSH keywords associated with a publication, in terms of the years since the first appearance of the MeSH keywords, to determine the level of novelty of that publication relative to the research frontier of the global scientific corpus. This approach, which considers the recency of knowledge elements in a scientific document, has been applied in various studies (Arts et al., 2021; Balsmeier et al., 2018; Fleming, 2001). Following this methodology, we compute the average age of all bio-entities extracted from a thesis to evaluate its novelty compared to the research frontier in bio-medical doctoral theses in the US. A higher average age indicates a less novel thesis. The results are outlined in Tables S10 to S12, and they generally align with the main findings.

## 6 Discussion and conclusion

### 6.1 Theoretical implications

This study offers significant theoretical contributions in multiple aspects. Firstly, it enriches the understanding of gender disparities in science by examining the lens of novel research strategies. Existing literature extensively discusses academic gender disparities through various performance metrics such as productivity (Eloy et al., 2013; Mayer & Rathmann, 2018), citation rates (Chatterjee & Werner, 2021;

Liu et al., 2020), funding opportunities (Larregue & Nielsen, 2023; Mirin, 2021), and prestigious awards (Ma et al., 2019). However, this body of literature rarely delves deeper into the underlying mechanisms that contribute to these gender differences in academic outcomes. This study complements previous research by providing empirical evidence of gender disparities in research strategies, specifically in the pursuit of novel research. The findings suggest that the notable gender disparities identified in generating novel research during the early-career stages may ultimately contribute to overall disadvantages faced by females in their academic careers.

The findings of this study highlight the importance of considering both institutional status and individual status when investigating gender disparities in science. Science is characterized by significant stratification (Davies & Zarifa, 2012; Merton, 1968), yet previous research on gender disparities often fails to account for potential variations in the magnitude of these disparities across different levels of university prestige and scientists' status (Fox, 2005; Larregue & Nielsen, 2023; Schaller, 2022). This study addresses these research gaps and enhances the policy relevance of its implications based on the obtained results. By incorporating these factors, a more comprehensive understanding of gender disparities in the field of science can be achieved.

In addition, this study contributes by presenting direct evidence of a decline in the novelty of knowledge at the individual level, specifically within doctoral research. The discourse on whether the innovation process is accelerating or decelerating is not a new topic (Bloom et al., 2020; Ellwood et al., 2017). However, most of these studies are conducted at the aggregate level, and yield conflicting conclusions (Jones, 2009; Liu, Bu, et al., 2022; Park et al., 2023). It remains uncertain whether generating novel research at the individual level is becoming more challenging or easier. This study unveils a clear downward trajectory in scientific novelty within doctoral theses, and this trend holds true across various levels of university prestige and genders.

## 6.2 Practical implications

This study makes important practical implications in two folds. First, the measurement of scientific novelty this study applies is an efficient and useful tool for the assessment of novel research publications and scientists who possess great potential to produce novel knowledge in biomedical disciplines or other similar disciplines. This study extends the application of the method (Liu, Bu, et al., 2022) that was originally proposed to measure coronavirus-related research articles to large-scale biomedical doctoral theses. Despite lacking ground truth data, this study applies another measurement of scientific novelty in life sciences, proposed by Azoulay et al. (2011), to confirm the reliability of the method to some extent. Building upon the perspective of combinatorial novelty, and using fine-grained measurements of knowledge units, the method to measure scientific novelty of doctoral theses offers a nuanced evaluation of early-career scientists' innovation potential to funding agencies, institutions managers and policymakers, which facilitates their decision-making processes regarding funding allocation, academic hiring, and promotions.

Furthermore, this study provides clear illustration of applying two advanced econometric regression models, i.e., fractional logistic and quantile regression models, to address research questions in information science. Fractional logistic regressions are used when the dependent variable is a continuous variable restricted within the bounded range of 0 to 1. This method is useful and provides precise estimations when we focus on metrics that are applied in information science research and range between 0 and 1, such as hit rates (Chen et al., 2023; Mele et al., 2020; Strotmann & Zhao, 2010), Gini coefficient (Xu et al., 2024), and some variations of disruptiveness score, i.e., CD-index (Chen et al., 2021;

Leydesdorff & Bornmann, 2021). Furthermore, this study applies quantile regression analyses that can be employed when we investigate heterogeneities concerning the distribution of dependent variables. The two advanced regression models have great potential to be applied in information science.

## 6.3 Conclusion

Based on 279,424 doctoral theses in biomedical sciences from US institutions from 1980 to 2016, this study explores how early-career scientists conducted novel research in their doctoral theses, gender disparities in this process, and the potential heterogeneities of such gender differences. To address the research questions, this study applies the perspective of combinatorial novelty and a pre-trained Bio-BERT model to measure scientific novelty in doctoral theses based on bio-entities, the basic knowledge units we investigate in this study.

This study reveals a declining trend in the scientific novelty of doctoral theses, which applies to different genders and affiliations with varying research intensities. This temporal decline is observed in both the fraction of novel theses and the average novelty scores. These findings suggest that, as knowledge and technology accumulate, the generation of innovative research becomes increasingly challenging, consistent with previous studies (Jones, 2010; Park et al., 2023). To address this growing difficulty, there are two potential strategies for junior scientists. Firstly, they can overcome limited research capacities resulting from educational burdens by collaborating with advisors and other researchers. Secondly, they can leverage emerging research techniques such as AI and LLMs to accelerate the innovative process (Wang et al., 2023).

Persistent and significant gender disparities have been observed in the scientific novelty of doctoral theses. Female students have a lower probability of producing a novel thesis, and their doctoral theses are less novel compared to those authored by men. These gender disparities also extend to the role of advisors. When supervised by female advisors, students tend to produce theses with a lower level of scientific novelty. Additionally, it is found that the supervision of female advisors exacerbates the gender disparities in the scientific novelty of theses. These findings align with a mechanism called the "female penalty of novelty" proposed by Trapido (2022). Novelty inherently involves uncertainties and risks and is more closely associated with the agentic orientation typically attributed to men rather than the communal orientation often assumed of women (Hora et al., 2021). Due to this gender stereotype, female scientists frequently receive less recognition for their novel contributions (Schmutz & Faupel, 2010; Taylor et al., 2020), which hampers their motivation to innovate. Moreover, the current research system lacks sufficient support for female scientists, failing to offset the hindering effects on their willingness to produce novel research. These obstacles are particularly pronounced for female students under the supervision of female advisors because both parties face the penalty of generating novel knowledge.

We further observe heterogeneities concerning gender disparities in the scientific novelty of doctoral theses. Specifically, such gender disparities are less prominent in top-tier universities. Research suggests that top-tier universities may be more supportive of female scientists, and provide a more friendly environment for them to innovate (Bothwell et al., 2022; Ceci & Williams, 2011), which may account for this finding. Furthermore, gender disparities in scientific novelty become increasingly slighter at the higher tails of the distribution of novelty scores, suggesting a smaller gender disparity of scientific novelty among female and male students who produced the highly novel doctoral research. This finding aligns with the high self-efficacy of high performers (Feather, 1989; McCoach & Siegle, 2001). When female students have the potential to become high performers, those who produced highly novel research in this case, may be more confident and tend to take more risks (Krueger Jr & Dickson, 1994; Macko & Tyszka, 2009), and thus produce highly novel research. Therefore, we observe slight gender differences in scientific novelty when we investigate the most highly novel theses. Taken together, those results suggest that the research environment and high self-confidence of students are critical for mitigating scientific disparities in producing novel research during doctoral study.

Scientific novelty plays a vital role in promoting scientific breakthroughs and technological innovation. Despite possessing similar creative skills and capabilities, female students face disadvantages in producing novel research within their doctoral theses. This deficiency in generating innovative research at the early stages of their careers can have long-term negative effects on the career development of female scientists and further widen gender disparities in science, particularly in terms of creating scientific breakthroughs later in their careers. To address these disparities, it is crucial to tackle systematic gender biases inherent in the current research system, particularly in evaluating novel contributions, and provide adequate institutional support for female faculty members and students. Additionally, funding agencies should offer more opportunities for funding to female scientists, encouraging them to pursue innovative research. Our findings indicate that the supervision of female advisors exacerbates the gender disparity in the scientific novelty of students' doctoral theses. This suggests that providing additional support to female faculty members not only directly benefits their career development but also holds significant importance in fostering the growth of future generations of female scientists.

This study has a few limitations. Due to the limitation of data availability, the study only focuses on the US, and the findings may not be applicable to other countries, especially countries with different research systems from the US. Second, this study only analyzes doctoral theses in biomedical science, not whether or not the findings hold for other disciplines, which should be investigated in future studies.

## Acknowledgements

This study is sponsored by the National Natural Science Foundation of China (72104054, 72104007), the Shanghai Pujiang Talent program (21PJC026), and the Key Project of the National Natural Science Foundation of China (72234001).

# Supplementary information

**Table S 1. Distribution of scientific domains of theses in PQDT.**

| Domain | Freq. | Percent | Cum. |
|---|---|---|---|
| Mathematical and Physical Sciences | 243,829 | 21.98 | 21.98 |
| Engineering | 206,485 | 18.61 | 40.59 |
| Biological Sciences | 201,018 | 18.12 | 58.71 |
| Behavioral Sciences | 158,358 | 14.27 | 72.98 |
| Health and Medical Sciences | 110,929 | 10 | 82.98 |
| Agriculture | 38,397 | 3.46 | 86.44 |
| Geosciences | 35,270 | 3.18 | 89.62 |
| Education | 26,781 | 2.41 | 92.03 |
| Social Sciences | 19,081 | 1.72 | 93.75 |
| Ecosystem Sciences | 14,712 | 1.33 | 95.08 |
| Interdisciplinary | 13,393 | 1.21 | 96.28 |
| Business | 10,668 | 0.96 | 97.24 |
| Environmental Sciences | 10,312 | 0.93 | 98.17 |
| Area, Ethnic, and Gender Studies | 6,764 | 0.61 | 98.78 |
| Philosophy and Religion | 4,332 | 0.39 | 99.17 |
| Communications and Information Sciences | 3,085 | 0.28 | 99.45 |
| Language & Literature | 2,316 | 0.21 | 99.66 |
| History | 1,298 | 0.12 | 99.78 |
| Fine and Performing Arts | 1,174 | 0.11 | 99.88 |
| Law and Legal Studies | 713 | 0.06 | 99.95 |
| Architecture | 573 | 0.05 | 100 |
| Language and Literature | 3 | 0 | 100 |
| Total | 1,109,491 | 100 | |

**Table S 2. Statistics of the original dataset.**

| Type of information | Number of observations |
|---|---|
| Doctoral theses | 279,424 |
| Years | 1980-2016 |
| Advisors | 114,083 |
| Universities | 708 |

**Table S 3. The prediction of students' and advisors' gender using Genderize.io.**

| | Male | Female | Female (%) | Total |
|---|---|---|---|---|
| **Student** | 121,267 | 115,257 | 48.73 | 236,524 |
| **Advisor** | 149,159 | 56,360 | 27.42 | 205,519 |

**Table S 4. The estimated gender disparities in the scientific novelty of doctoral theses when including data on observations with the gender prediction of "Mostly female" and "Mostly male".**

| | (1) | (2) | (3) | (1) | (2) | (3) |
|---|---|---|---|---|---|---|
| Variable | Novel combination | | | Novelty score | | |
| **Female student** | -0.286*** | -0.213*** | -0.089*** | -0.082*** | -0.066*** | -0.043*** |
| | (0.010) | (0.012) | (0.013) | (0.008) | (0.009) | (0.010) |
| **Female advisor** | | -0.416*** | -0.117*** | | -0.082*** | -0.021 |
| | | (0.013) | (0.021) | | (0.011) | (0.016) |
| **Female student × Female advisor** | | | -0.481*** | | | -0.100*** |
| | | | (0.027) | | | (0.022) |
| **Controls** | YES | YES | YES | YES | YES | YES |

| Year FE | YES | YES | YES | YES | YES | YES |
|---|---|---|---|---|---|---|
| Observations | 159,813 | 135,683 | 135,683 | 159,814 | 135,684 | 135,684 |
| pseudo $R^2$ | 0.0410 | 0.0469 | 0.0487 | 0.0019 | 0.0021 | 0.0021 |

Notes: Robust seeform are in parentheses. *** p<0.01, ** p<0.05, * p<0.1.

Table S 5. The estimated gender disparities in the scientific novelty of doctoral theses across prestige of university when including data on observations with the gender prediction of "Mostly female" and "Mostly male".

|  | (1) | (2) | (3) | (4) |
|---|---|---|---|---|
| Variable | Novel combination | | Novelty score | |
|  | R1 university | Non-R1 university | R1 university | Non-R1 university |
| Female student | -0.194*** | -0.511*** | -0.055*** | -0.151*** |
|  | (0.012) | (0.020) | (0.009) | (0.017) |
| Controls | YES | YES | YES | YES |
| Year FE | YES | YES | YES | YES |
| Observations | 117,276 | 44,670 | 117,280 | 44,671 |
| pseudo $R^2$ | 0.0326 | 0.0627 | 0.0010 | 0.0049 |

Notes: Robust seeform are in parentheses. *** p<0.01, ** p<0.05, * p<0.1.

Table S 6. The estimated gender disparities in the scientific novelty of doctoral theses across different percentiles of scientific novelty when including data on observations with the gender prediction of "Mostly female" and "Mostly male".

|  | (1) | (2) | (3) | (4) | (5) |
|---|---|---|---|---|---|
| Variable | Q50 | Q60 | Q70 | Q80 | Q90 |
| Female student | -0.008*** | -0.012*** | -0.010*** | -0.006*** | -0.004*** |
|  | (0.000) | (0.001) | (0.001) | (0.001) | (0.001) |
| Controls | YES | YES | YES | YES | YES |
| Year Fixed Effect | Yes | Yes | Yes | Yes | Yes |
| Observations | 137,295 | 137,295 | 137,295 | 137,295 | 137,295 |
| pseudo $R^2$ | 0.0333 | 0.0221 | 0.0103 | 0.00413 | 0.00273 |

Notes: Robust seeform are in parentheses. *** p<0.01, ** p<0.05, * p<0.1.

Table S 7. The estimated gender disparities in the scientific novelty of doctoral theses based on the predicted gender information from Genderize.io.

|  | (1) | (2) | (3) | (4) | (5) | (6) |
|---|---|---|---|---|---|---|
| Variable | Novel combination | | | Novelty score | | |
| Female student | -0.312*** | -0.237*** | -0.103*** | -0.080*** | -0.065*** | -0.044*** |
|  | (0.010) | (0.010) | (0.012) | (0.007) | (0.008) | (0.009) |
| Female advisor |  | -0.437*** | -0.119*** |  | -0.090*** | -0.038*** |
|  |  | (0.011) | (0.019) |  | (0.010) | (0.014) |
| Female student × Female advisor |  |  | -0.510*** |  |  | -0.086*** |
|  |  |  | (0.024) |  |  | (0.019) |
| Controls | YES | YES | YES | YES | YES | YES |
| Year FE | YES | YES | YES | YES | YES | YES |
| Observations | 189,308 | 169,788 | 169,788 | 189,309 | 169,789 | 169,789 |
| pseudo $R^2$ | 0.0405 | 0.0473 | 0.0493 | 0.0017 | 0.0019 | 0.0020 |

Notes: Robust seeform are in parentheses. *** p<0.01, ** p<0.05, * p<0.1.

Table S 8. The estimated gender disparities in the scientific novelty of doctoral theses across prestige of university based on the predicted gender information from Genderize.io.

|  | (1) | (2) | (3) | (4) |
|---|---|---|---|---|
| Variable | Novel combination | | Novelty score | |
|  | R1 university | Non-R1 university | R1 university | Non-R1 university |
| Female student | -0.211*** | -0.559*** | -0.054*** | -0.145*** |
|  | (0.011) | (0.019) | (0.008) | (0.015) |
| Controls | YES | YES | YES | YES |
| Year FE | YES | YES | YES | YES |
| Observations | 139,127 | 52,772 | 139,131 | 52,773 |
| pseudo $R^2$ | 0.0324 | 0.0641 | 0.0009 | 0.0045 |

Notes: Robust seeform are in parentheses. *** p<0.01, ** p<0.05, * p<0.1.

Table S 9. The estimated gender disparities in the scientific novelty of doctoral theses across different percentiles of scientific novelty based on the predicted gender information from Genderize.io.

|  | (1) | (2) | (3) | (4) | (5) |
|---|---|---|---|---|---|
| Variable | Q50 | Q60 | Q70 | Q80 | Q90 |
| Female student | -0.010*** | -0.013*** | -0.010*** | -0.006*** | -0.003*** |
|  | (0.000) | (0.001) | (0.001) | (0.001) | (0.001) |
| Controls | YES | YES | YES | YES | YES |
| Year Fixed Effect | Yes | Yes | Yes | Yes | Yes |
| Observations | 171,832 | 171,832 | 171,832 | 171,832 | 171,832 |
| pseudo $R^2$ | 0.0324 | 0.0210 | 0.00949 | 0.00386 | 0.00247 |

Notes: Robust seeform are in parentheses. *** p<0.01, ** p<0.05, * p<0.1.

Table S 10. The estimated gender disparities in the scientific novelty of doctoral theses when using the average mean age of bio-entities to measure scientific novelty.

| Variable | (1) | (2) | (3) |
|---|---|---|---|
| Female student | 0.074*** | 0.061*** | 0.034*** |
|  | (0.002) | (0.002) | (0.002) |
| Female advisor |  | 0.094*** | 0.029*** |
|  |  | (0.002) | (0.003) |
| Female student × female advisor |  |  | 0.103*** |
|  |  |  | (0.004) |
| Controls | YES | YES | YES |
| Year FE | YES | YES | YES |
| Observations | 173,535 | 155,149 | 155,149 |
| R-squared | 0.251 | 0.263 | 0.267 |

Notes: Robust standard errors are in parentheses. *** p<0.01, ** p<0.05, * p<0.1.

Table S 11. The estimated gender disparities in the scientific novelty of doctoral theses across prestige of university when using the average mean age of bio-entities to measure scientific novelty.

| Variable | (1) | (2) |
|---|---|---|
| Female student | 0.057*** | 0.119*** |
|  | (0.002) | (0.003) |
| Controls | YES | YES |
| Year FE | YES | YES |
| Observations | 127,207 | 48,624 |
| R-squared | 0.234 | 0.278 |

**Notes**: Robust standard errors are in parentheses. *** p<0.01, ** p<0.05, * p<0.1.

**Table S 12. The estimated gender disparities in the scientific novelty of doctoral theses across different percentiles of scientific novelty when using the average mean age of bio-entities to measure scientific novelty.**

|  | (1) | (2) | (3) | (4) | (5) |
|---|---|---|---|---|---|
|  | **Q50** | **Q60** | **Q70** | **Q80** | **Q90** |
| **Female student** | 0.058*** | 0.062*** | 0.069*** | 0.072*** | 0.072*** |
|  | (0.002) | (0.002) | (0.002) | (0.003) | (0.003) |
| **Controls** | YES | YES | YES | **Controls** | YES |
| **Year Fixed Effect** | Yes | Yes | Yes | Yes | Yes |
| **Observations** | 156,979 | 156,979 | 156,979 | 156,979 | 156,979 |
| **pseudo $R^2$** | 0.152 | 0.147 | 0.140 | 0.129 | 0.111 |

**Notes**: Robust standard errors are in parentheses. *** p<0.01, ** p<0.05, * p<0.1.

# References

Aguinis, H., Ji, Y. H., & Joo, H. (2018). Gender productivity gap among star performers in STEM and other scientific fields. *Journal of Applied Psychology*, *103*(12), 1283.

Allison, P. D., & Long, J. S. (1990). Departmental effects on scientific productivity. *American Sociological Review*, 469-478.

Arts, S., Hou, J., & Gomez, J. C. (2021). Natural language processing to identify the creation and impact of new technologies in patent text: Code, data, and new measures. *Research Policy*, *50*(2), 104144.

Azoulay, P., Graff Zivin, J. S., & Manso, G. (2011). Incentives and creativity: evidence from the academic life sciences. *The RAND Journal of Economics*, *42*(3), 527-554.

Baer, J., & Kaufman, J. C. (2008). Gender differences in creativity. *The Journal of Creative Behavior*, *42*(2), 75-105.

Balsmeier, B., Assaf, M., Chesebro, T., Fierro, G., Johnson, K., Johnson, S., Li, G. C., Lück, S., O'Reagan, D., & Yeh, B. (2018). Machine learning and natural language processing on the patent corpus: Data, tools, and new measures. *Journal of Economics & Management Strategy*, *27*(3), 535-553.

Benson, A., Li, D., & Shue, K. (2021). "Potential" and the gender promotion gap. *Unpublished Working Paper*. https://danielle-li.github.io/assets/docs/PotentialAndTheGenderPromotionGap.pdf

Bloom, N., Jones, C. I., Van Reenen, J., & Webb, M. (2020). Are ideas getting harder to find? *American Economic Review*, *110*(4), 1104-1144.

Bornmann, L., Mutz, R., & Daniel, H.-D. (2007). Gender differences in grant peer review: A meta-analysis. *Journal of Informetrics*, *1*(3), 226-238.

Borsuk, R. M., Aarssen, L. W., Budden, A. E., Koricheva, J., Leimu, R., Tregenza, T., & Lortie, C. J. (2009). To name or not to name: The effect of changing author gender on peer review. *BioScience*, *59*(11), 985-989.

Bothwell, E., Roser Chinchilla, J. F., Deraze, E., Ellis, R., Galán Muros, V., Gallegos, G., & Mutize, T. (2022). *Gender equality: How global universities are performing*. https://redined.educacion.gob.es/xmlui/handle/11162/240963

Boudreau, K. J., Guinan, E. C., Lakhani, K. R., & Riedl, C. (2016). Looking across and looking beyond the knowledge frontier: Intellectual distance, novelty, and resource allocation in science. *Management Science*, *62*(10), 2765-2783.

Breda, T., Grenet, J., Monnet, M., & Van Effenterre, C. (2023). How Effective are Female Role Models in Steering Girls Towards Stem? Evidence from French High Schools. *The Economic Journal*, *133*(653), 1773-1809.

Budden, A. E., Tregenza, T., Aarssen, L. W., Koricheva, J., Leimu, R., & Lortie, C. J. (2008). Double-blind review favours increased representation of female authors. *Trends in Ecology & Evolution*, *23*(1), 4-6.

Carli, L. L., Alawa, L., Lee, Y., Zhao, B., & Kim, E. (2016). Stereotypes about gender and science: Women ≠ scientists. *Psychology of Women Quarterly*, *40*(2), 244-260.

Ceci, S. J., & Williams, W. M. (2011). Understanding current causes of women's underrepresentation in science. *Proceedings of the National Academy of Sciences*, *108*(8),

3157-3162.

Chai, S., & Menon, A. (2019). Breakthrough recognition: Bias against novelty and competition for attention. *Research Policy*, *48*(3), 733-747.

Chan, H. F., & Torgler, B. (2020). Gender differences in performance of top cited scientists by field and country. *Scientometrics*, *125*(3), 2421-2447.

Chavez-Eakle, R. A., Lara, M. d. C., & Cruz-Fuentes, C. (2006). Personality: A possible bridge between creativity and psychopathology? *Creativity Research Journal*, *18*(1), 27-38.

Chen, J., Shao, D., & Fan, S. (2021). Destabilization and consolidation: Conceptualizing, measuring, and validating the dual characteristics of technology. *Research Policy*, *50*(1), 104115. https://doi.org/https://doi.org/10.1016/j.respol.2020.104115

Chen, R., Fan, J., & Wu, M. (2023). MC-RGN: Residual Graph Neural Networks based on Markov Chain for sequential recommendation. *Information Processing & Management*, *60*(6), 103519.

Chiu, B., & Baker, S. (2020). Word embeddings for biomedical natural language processing: A survey. *Language and Linguistics Compass*, *14*(12), e12402.

Cockburn, I. M., Henderson, R., & Stern, S. (2018). The impact of artificial intelligence on innovation: An exploratory analysis. In *The economics of artificial intelligence: An agenda* (pp. 115-146). University of Chicago Press.

Cohen, B. A. (2017). How should novelty be valued in science? *Elife*, *6*, e28699.

Council, N. R. (2010). *Gender differences at critical transitions in the careers of science, engineering, and mathematics faculty*. National Academies Press.

Criscuolo, P., Dahlander, L., Grohsjean, T., & Salter, A. (2017). Evaluating novelty: The role of panels in the selection of R&D projects. *Academy of Management Journal*, *60*(2), 433-460.

D'este, P., Llopis, O., Rentocchini, F., & Yegros, A. (2019). The relationship between interdisciplinarity and distinct modes of university-industry interaction. *Research Policy*, *48*(9), 103799.

Davies, S., & Zarifa, D. (2012). The stratification of universities: Structural inequality in Canada and the United States. *Research in Social Stratification and Mobility*, *30*(2), 143-158.

Dul, J., Ceylan, C., & Jaspers, F. (2011). Knowledge workers' creativity and the role of the physical work environment. *Human Resource Management*, *50*(6), 715-734.

Dwivedi, Y. K., Kshetri, N., Hughes, L., Slade, E. L., Jeyaraj, A., Kar, A. K., Baabdullah, A. M., Koohang, A., Raghavan, V., & Ahuja, M. (2023). "So what if ChatGPT wrote it?" Multidisciplinary perspectives on opportunities, challenges and implications of generative conversational AI for research, practice and policy. *International Journal of Information Management*, *71*, 102642.

Eagly, A. H., & Karau, S. J. (2002). Role congruity theory of prejudice toward female leaders. *Psychological review*, *109*(3), 573.

Eagly, A. H., Nater, C., Miller, D. I., Kaufmann, M., & Sczesny, S. (2020). Gender stereotypes have changed: A cross-temporal meta-analysis of US public opinion polls from 1946 to 2018. *American Psychologist*, *75*(3), 301.

Ellwood, P., Grimshaw, P., & Pandza, K. (2017). Accelerating the innovation process: A systematic review and realist synthesis of the research literature. *International Journal of Management Reviews*, *19*(4), 510-530.

Eloy, J. A., Svider, P. F., Cherla, D. V., Diaz, L., Kovalerchik, O., Mauro, K. M., Baredes, S., &

Chandrasekhar, S. S. (2013). Gender disparities in research productivity among 9952 academic physicians. *The Laryngoscope*, *123*(8), 1865-1875.

Feather, N. T. (1989). Attitudes towards the high achiever: The fall of the tall poppy. *Australian Journal of Psychology*, *41*(3), 239-267.

Fleming, L. (2001). Recombinant uncertainty in technological search. *Management Science*, *47*(1), 117-132.

Fontana, M., Iori, M., Montobbio, F., & Sinatra, R. (2020). New and atypical combinations: An assessment of novelty and interdisciplinarity. *Research Policy*, *49*(7), 104063.

Foster, J. G., Rzhetsky, A., & Evans, J. A. (2015). Tradition and innovation in scientists' research strategies. *American Sociological Review*, *80*(5), 875-908.

Fox, M. F. (2005). Gender, family characteristics, and publication productivity among scientists. *Social studies of science*, *35*(1), 131-150.

Goldin, C., & Rouse, C. (2000). Orchestrating impartiality: The impact of "blind" auditions on female musicians. *American Economic Review*, *90*(4), 715-741.

Hampole, M., Truffa, F., & Wong, A. (2021). *Peer effects and the gender gap in corporate leadership: Evidence from MBA students*. https://fass.nus.edu.sg/ecs/wp-content/uploads/sites/4/2022/09/Peer-Effects-and-the-Gender-Gap-in-Corporate-Leadership-Evidence-from-MBA-Students.pdf

Holman, L., Stuart-Fox, D., & Hauser, C. E. (2018). The gender gap in science: How long until women are equally represented? *PLoS biology*, *16*(4), e2004956.

Hopkins, N. (2002). A study on the status of women faculty in science at MIT. AIP Conference Proceedings,

Hora, S., Lemoine, G. J., Xu, N., & Shalley, C. E. (2021). Unlocking and closing the gender gap in creative performance: A multilevel model. *Journal of Organizational Behavior*, *42*(3), 297-312.

Huang, J., Gates, A. J., Sinatra, R., & Barabási, A.-L. (2020). Historical comparison of gender inequality in scientific careers across countries and disciplines. *Proceedings of the National Academy of Sciences*, *117*(9), 4609-4616.

Jacobs, J. A. (1996). Gender inequality and higher education. *Annual Review of Sociology*, *22*(1), 153-185.

Jones, B. F. (2009). The burden of knowledge and the "death of the renaissance man": Is innovation getting harder? *The Review of Economic Studies*, *76*(1), 283-317.

Jones, B. F. (2010). Age and great invention. *The Review of Economics and Statistics*, *92*(1), 1-14.

Kabat-Farr, D., & Cortina, L. M. (2012). Selective incivility: Gender, race, and the discriminatory workplace. In T. R. L. Suzy Fox (Ed.), *Gender and the dysfunctional workplace* (pp. 120-134). Edward Elgar Publishing.

Karami, A., White, C. N., Ford, K., Swan, S., & Spinel, M. Y. (2020). Unwanted advances in higher education: Uncovering sexual harassment experiences in academia with text mining. *Information Processing & Management*, *57*(2), 102167.

Kim, D., Lee, J., So, C. H., Jeon, H., Jeong, M., Choi, Y., Yoon, W., Sung, M., & Kang, J. (2019). A neural named entity recognition and multi-type normalization tool for biomedical text mining. *IEEE Access*, *7*, 73729-73740.

Kittur, A., Yu, L., Hope, T., Chan, J., Lifshitz-Assaf, H., Gilon, K., Ng, F., Kraut, R. E., & Shahaf, D. (2019). Scaling up analogical innovation with crowds and AI. *Proceedings of the National*

*Academy of Sciences*, *116*(6), 1870-1877.

Koenker, R., & Bassett Jr, G. (1978). Regression quantiles. *Econometrica: journal of the Econometric Society*, 33-50.

Koenker, R., & Hallock, K. F. (2001). Quantile regression. *Journal of economic perspectives*, *15*(4), 143-156.

Kogan, N. (1974). Creativity and Sex Differences. *The Journal of Creative Behavior*, *8*(1), 1-14.

Krueger Jr, N., & Dickson, P. R. (1994). How believing in ourselves increases risk taking: Perceived self-efficacy and opportunity recognition. *Decision Sciences*, *25*(3), 385-400.

Larivière, V., Vignola-Gagné, E., Villeneuve, C., Gélinas, P., & Gingras, Y. (2011). Sex differences in research funding, productivity and impact: an analysis of Québec university professors. *Scientometrics*, *87*(3), 483-498.

Larregue, J., & Nielsen, M. W. (2023). Knowledge Hierarchies and Gender Disparities in Social Science Funding. *Sociology*, 00380385231163071.

Leydesdorff, L., & Bornmann, L. (2021). Disruption indices and their calculation using web-of-science data: Indicators of historical developments or evolutionary dynamics? *Journal of Informetrics*, *15*(4), 101219.

Liang, Z., Mao, J., & Li, G. (2023). Bias against scientific novelty: A prepublication perspective. *Journal of the Association for Information Science and Technology*, *74*(1), 99-114.

Liao, Z., Zhang, J. H., WANG, N., Bottom, W. P., Deichmann, D., & Tang, P. M. (2023). The Gendered Liability of Venture Novelty. *Academy of Management Journal*(ja).

Liu, F., Holme, P., Chiesa, M., AlShebli, B., & Rahwan, T. (2023). Gender inequality and self-publication are common among academic editors. *Nature Human Behaviour*, *7*(3), 353-364.

Liu, M., Bu, Y., Chen, C., Xu, J., Li, D., Leng, Y., Freeman, R. B., Meyer, E. T., Yoon, W., & Sung, M. (2022). Pandemics are catalysts of scientific novelty: Evidence from COVID-19. *Journal of the Association for Information Science and Technology*, *73*(8), 1065-1078.

Liu, M., & Hu, X. (2022). Movers' advantages: The effect of mobility on scientists' productivity and collaboration. *Journal of Informetrics*, *16*(3), 101311.

Liu, M., Zhang, N., Hu, X., Jaiswal, A., Xu, J., Chen, H., Ding, Y., & Bu, Y. (2022). Further divided gender gaps in research productivity and collaboration during the COVID-19 pandemic: Evidence from coronavirus-related literature [Article]. *Journal of Informetrics*, *16*(2), Article 101295. https://doi.org/10.1016/j.joi.2022.101295

Lubienski, S. T., Miller, E. K., & Saclarides, E. S. (2018). Sex differences in doctoral student publication rates. *Educational Researcher*, *47*(1), 76-81.

Luksyte, A., Unsworth, K. L., & Avery, D. R. (2018). Innovative work behavior and sex-based stereotypes: Examining sex differences in perceptions and evaluations of innovative work behavior. *Journal of Organizational Behavior*, *39*(3), 292-305.

Ma, Y., Oliveira, D. F., Woodruff, T. K., & Uzzi, B. (2019). Women who win prizes get less money and prestige. *Nature*, *565*(7739), 287-288.

Macko, A., & Tyszka, T. (2009). Entrepreneurship and risk taking. *Applied Psychology*, *58*(3), 469-487.

Martín-Brufau, R., & Corbalán, J. (2016). Creativity and psychopathology: Sex matters. *Creativity Research Journal*, *28*(2), 222-228.

Mayer, S. J., & Rathmann, J. M. (2018). How does research productivity relate to gender? Analyzing gender differences for multiple publication dimensions. *Scientometrics*, *117*(3), 1663-1693.

McCoach, D. B., & Siegle, D. (2001). A comparison of high achievers' and low achievers' attitudes, perceptions, and motivations. *Academic Exchange*, *2*, 71-76.

Mele, I., Tonellotto, N., Frieder, O., & Perego, R. (2020). Topical result caching in web search engines. *Information Processing & Management*, *57*(3), 102193.

Merton, R. K. (1968). The Matthew effect in science: The reward and communication systems of science are considered. *Science*, *159*(3810), 56-63.

Mirin, A. A. (2021). Gender disparity in the funding of diseases by the US National Institutes of Health. *Journal of Women's Health*, *30*(7), 956-963.

Nittrouer, C. L., Hebl, M. R., Ashburn-Nardo, L., Trump-Steele, R. C., Lane, D. M., & Valian, V. (2018). Gender disparities in colloquium speakers at top universities. *Proceedings of the National Academy of Sciences*, *115*(1), 104-108.

Park, M., Leahey, E., & Funk, R. J. (2023). Papers and patents are becoming less disruptive over time. *Nature*, *613*(7942), 138-144.

Reuben, E., Sapienza, P., & Zingales, L. (2014). How stereotypes impair women's careers in science. *Proceedings of the National Academy of Sciences*, *111*(12), 4403-4408.

Rossiter, M. W. (1993). The Matthew Matilda effect in science. *Social studies of science*, *23*(2), 325-341.

Santamaría, L., & Mihaljević, H. (2018). Comparison and benchmark of name-to-gender inference services. *PeerJ Computer Science*, *4*, e156.

Schaller, M. D. (2022). The gender gap amongst doctoral students in the biomedical sciences. *BioRXiv*, 2022.2010. 2018.512765.

Schmutz, V., & Faupel, A. (2010). Gender and cultural consecration in popular music. *Social forces*, *89*(2), 685-707.

Schumpeter, J. A. (1939). *Business cycles* (Vol. 1). Mcgraw-hill New York.

Sebo, P. (2021). Using genderize. io to infer the gender of first names: how to improve the accuracy of the inference. *Journal of the Medical Library Association: JMLA*, *109*(4), 609.

Sheltzer, J. M., & Smith, J. C. (2014). Elite male faculty in the life sciences employ fewer women. *Proceedings of the National Academy of Sciences*, *111*(28), 10107-10112.

Shen, H. (2013). Inequality quantified: Mind the gender gap. *Nature News*, *495*(7439), 22.

Simonton, D. K. (2003). Scientific creativity as constrained stochastic behavior: the integration of product, person, and process perspectives. *Psychological Bulletin*, *129*(4), 475.

Squazzoni, F., Bravo, G., Farjam, M., Marusic, A., Mehmani, B., Willis, M., Birukou, A., Dondio, P., & Grimaldo, F. (2021). Peer review and gender bias: A study on 145 scholarly journals. *Science Advances*, *7*(2), eabd0299.

Strotmann, A., & Zhao, D. (2010). Combining commercial citation indexes and open-access bibliographic databases to delimit highly interdisciplinary research fields for citation analysis. *Journal of Informetrics*, *4*(2), 194-200.

Sung, M., Jeong, M., Choi, Y., Kim, D., Lee, J., & Kang, J. (2022). BERN2: an advanced neural biomedical named entity recognition and normalization tool. *Bioinformatics*, *38*(20), 4837-4839.

Taylor, C., Ivcevic, Z., Moeller, J., & Brackett, M. (2020). Gender and support for creativity at work. *Creativity and Innovation Management*, *29*(3), 453-464.

Trapido, D. (2015). How novelty in knowledge earns recognition: The role of consistent identities. *Research Policy*, *44*(8), 1488-1500.

Trapido, D. (2022). The female penalty for novelty and the offsetting effect of alternate status characteristics. *Social Forces*, *100*(4), 1592-1618.

Uzzi, B., Mukherjee, S., Stringer, M., & Jones, B. (2013). Atypical combinations and scientific impact. *Science*, *342*(6157), 468-472.

Wang, D. Q., Feng, L. Y., Ye, J. G., Zou, J. G., & Zheng, Y. F. (2023). Accelerating the integration of ChatGPT and other large-scale AI models into biomedical research and healthcare. *MedComm–Future Medicine*, *2*(2), e43.

Wang, J., & Shibayama, S. (2022). Mentorship and creativity: Effects of mentor creativity and mentoring style. *Research Policy*, *51*(3), 104451.

Wang, J., Veugelers, R., & Stephan, P. (2017). Bias against novelty in science: A cautionary tale for users of bibliometric indicators. *Research Policy*, *46*(8), 1416-1436.

Way, S. F., Larremore, D. B., & Clauset, A. (2016). Gender, productivity, and prestige in computer science faculty hiring networks. Proceedings of the 25th international conference on world wide web,

Weitzman, M. L. (1998). Recombinant growth. *The Quarterly Journal of Economics*, *113*(2), 331-360.

Wuestman, M., Wanzenböck, I., & Frenken, K. (2023). Local peer communities and future academic success of Ph. D. candidates. *Research Policy*, *52*(8), 104844.

Xing, Y., Fan, Y., Sinatra, R., & Zeng, A. (2022). Academic mentees succeed in big groups, but thrive in small groups. *arXiv preprint arXiv:2208.05304*.

Xu, H., Liu, M., Bu, Y., Sun, S., Zhang, Y., Zhang, C., Acuna, D. E., Gray, S., Meyer, E., & Ding, Y. (2024). The impact of heterogeneous shared leadership in scientific teams. *Information Processing & Management*, *61*(1), 103542.

Yin, D., Wu, Z., Yokota, K., Matsumoto, K., & Shibayama, S. (2023). Identify novel elements of knowledge with word embedding. *Plos One*, *18*(6), e0284567.

Zenger, J., & Folkman, J. (2019). Women score higher than men in most leadership skills. *Harvard Business Review*, *92*(10), 86-93.

Zhang, L., Shang, Y., Huang, Y., & Sivertsen, G. (2022). Gender differences among active reviewers: an investigation based on Publons. *Scientometrics*, *127*(1), 145-179.
**35 / 35**